\documentclass{article}
\usepackage[utf8]{inputenc}
\usepackage{graphicx}
\usepackage{amsmath}
\usepackage{xcolor}
\newtheorem{definition}{Definition}
\newtheorem{theorem}{Theorem}

\newtheorem{proposition}{Proposition}
\newtheorem{remark}{Remark}
\usepackage{subcaption}
\usepackage{hyperref}
 \usepackage{setspace}
\usepackage[toc,page,titletoc]{appendix}

\providecommand{\keywords}[1]
{
  \small	
  \textbf{\textit{Keywords}} #1
}
\title{Variance of entropy for testing time-varying regimes with an application to meme stocks}
\author{Andrey Shternshis$^{(1)}$, Piero Mazzarisi$^{(1,2)}$\\
\\
        \small $^{(1)}$Scuola Normale Superiore, Piazza dei Cavalieri 7, Pisa, Italy, 56126\\
        \small $^{(2)}$University of Siena, Department of Economics and Statistics,\\ \small Piazza S.Francesco, 7-8, Siena, Italy, 53100}

\begin{document}

\maketitle
\begin{abstract}
Shannon entropy is the most common metric to measure the degree of randomness of time series in many fields, ranging from physics and finance to medicine and biology. Real-world systems may be in general non stationary, with an entropy value that is not constant in time. 
The goal of this paper is to propose a hypothesis testing procedure to test the null hypothesis of constant Shannon entropy for time series, against the alternative of a significant variation of the entropy between two subsequent periods. To this end, we find an unbiased approximation of the variance of the Shannon entropy’s estimator, up to the order $O(n^{-4})$ with $n$ the sample size. In order to characterize the variance of the estimator, we first obtain the explicit formulas of the central moments for both the binomial and the multinomial distributions, which describe the distribution of the Shannon entropy. Second, we find the optimal length of the rolling window used for estimating the time-varying Shannon entropy by optimizing a novel self-consistent criterion based on the counting of significant variations of entropy within a time window. We corroborate our findings by using the novel methodology to test for time-varying regimes of entropy for stock price dynamics, in particular considering the case of meme stocks in 2020 and 2021. We empirically show the existence of periods of market inefficiency for meme stocks. In particular, sharp increases of prices and trading volumes correspond to statistically significant drops of Shannon entropy.
\end{abstract}
\keywords{Entropy distribution, approximate computation of entropy variance, hypothesis testing for market efficiency, meme stocks}

\section{Introduction}
The Shannon entropy is widely used as a measure of randomness in many fields, such as finance, physics, medicine, and biology \cite{Pincus91, Pincus04, Dong19, Pandey15, Strait96, Bezerianos03}. One of the main applications of entropy estimation in finance is to measure the randomness of price returns. When the price incorporates all relevant information, the market is called efficient and the price dynamics is a martingale \cite{Samuelson, Fama}. As such, the Shannon entropy takes the maximum allowed value and can then be used as a measure of market efficiency. For the review of methods for testing martingale property and nonlinear dynamics in financial time series, we refer to the article \cite{Barnett00}. The drivers of market dynamics are however the result of the complex process of matching the supply and the demand of a large number of investors. It is easy to imagine that the market does not necessarily reflect all relevant information at certain times, because of the complex nature of the price formation mechanism. Moreover, feedback loops, irrational agents, market panic and speculation, or coordination of retail investors driven by non-economic reasons (like for the GameStop case) are just a few examples of mechanisms which can potentially create booms and busts \cite{Agliari18, Chan21}, thus driving the price far away from its fundamental value. In such cases, the price dynamics may display some level of predictability and, as such, the market is inefficient. This can be captured by some low value of the Shannon entropy.

In a number of works, the hypothesis of market efficiency is relaxed, by accounting for the possibility of periods of inefficiencies. In order to capture such an effect, Shannon entropy is considered as time-varying and it is computed by using a window rolling over the period of interest, see, e.g., \cite{Molgedey00, Risso08, Mensi,Olbrys}. It is important to notice that many patterns of price dynamics may jeopardize the estimate of entropy. For example, long memory of volatility can be the result of a regime switching behavior  \cite{susmel2000, Lobo98, Malik05} and, as a consequence, the estimate of any other dynamic pattern can be affected. For this reason, it is important to filter out any known pattern of market predictability, such as heteroscedasticity or seasonality, before using the estimate of entropy as measure of market efficiency. Interestingly, even after such filtering, the price dynamics often continues to display some predictability that is captured by low values of entropy estimates, signaling the not complete efficiency of the market \cite{Calcagnile,Shternshis2}. A crucial question is whether a drop in the entropy estimate at some period is statistically significant or it is only a fluctuation consistent with the null hypothesis of market efficiency.

The goal of this paper is to propose a rigorous methodology to identify significant changes in the value of the Shannon entropy. 
We define a rigorous procedure to test the hypothesis that the entropy value associated with two different sequences (defined on the same finite alphabet) is statistically the same. To this end, we need to solve two problems. The first problem is \textit{to find the variance of the Shannon entropy’s estimator} obtained by the Empirical Frequencies method \cite{Marton}. The second problem is \textit{to find the optimal length of a rolling window} used to estimate the time-varying entropy of a time series.

The variance of the Shannon entropy is the key quantity to use in order to determine if two entropy estimates associated with two different sequences are statistically equal. In fact, it is possible to define a z-score given the variance of the Shannon entropy, thus determining the corresponding p-value of statistically equal estimates.

G.~P.~Basharin \cite{basharin59} obtained the first order approximation of the variance of the entropy estimator $\hat{H}$ calculated using the Empirical Frequencies method. 
\begin{equation}
\label{eq:D1}
D_1(\hat{H})=\frac{1}{n}(\sum_j p_j\ln^2{p_j}-H^2)
\end{equation}

where $p_j$ are the set of the probabilities of possible events and $n$ is the length of the sequence of events. The same result was later obtained in \cite{Jabloun}. However, Eq.~\ref{eq:D1} holds in the asymptotic regime, that is when the length of the sequence becomes arbitrarily large, i.e. $n\rightarrow\infty$. Here, we aim to estimate the time-varying entropy for finite samples, i.e. using a finite length $n$ of the sequence. Moreover, Eq.~\ref{eq:D1} is not a consistent estimator of the variance in the case of equal probabilities. When all probabilities $p_j$ are equal, $D_1(\hat{H})=0$. Thus, a more accurate approximation for the variance is needed.\footnote{For example, B.~Harris \cite{Harris75} found the approximation with an error term $O(n^{-3})$.}
 
Here, we obtain a formula for the variance of the estimator of the Shannon entropy as a sum of central moments of binomial and multinomial distributions. The central moments are calculated by a new recursive approach. This helps us to find the approximation of the variance with an accuracy of order $O(n^{-4})$.\footnote{In general, it is possible to further extend such an approximation by using the proposed approach to compute higher orders of the central moments associated with the multinomial distribution.}

Interestingly, Ricci et al. \cite{Ricci21} found that the naive estimator of the variance approximation $D_1(\hat{H})$ in Eq.~\ref{eq:D1} has a bias term of order $O(n^{-2})$. On the contrary, we show that our proposed estimation for the variance of entropy is unbiased. Finally, by leveraging on the explicit formulation of the variance of the estimator of entropy, we are able to define a statistical test for entropy variation. In \cite{MATILLAGARCIA07}, the author suggests a statistical test for independence of symbolic dynamics. The test statistics is related to the permutation entropy. In our research, we are not restricted to the case when the benchmark value of the entropy is its maximum. A rejection of the null hypothesis signals statistically significant variation between any two possible values of entropy.

The second problem relies on finding the optimal length, i.e. the bandwidth, of the time window when testing for entropy variation between two subsequent time series. This is a problem of bias-variance tradeoff: reducing the length of the window allows to obtain a timely estimate of entropy, i.e. small bias, at the expense of increasing the variance of the estimator, and vice versa. 

Here, we introduce a novel self-consistent criterion to select (in-sample) the optimal bandwidth $\omega$. Given a sample of size $T$, we first define a counting function of the percentage of entropy variations for non-overlapping time series of length $\omega$ within such sample. Under the assumption of a finite number of {\it true} entropy variations, the percentage of estimated variations becomes negligible when $\omega$ is close to the minimum ($\omega=1$), while it is zero by definition when $\omega$ attains its maximum ($\omega=T$). Then, we show by simulations that such counting function has one maximum corresponding to the optimal bandwidth.


We use the novel methodology to find significant changes in entropy on simulated and real data. We investigate changes in the efficiency of the New York Stock Exchange with a special focus on meme stocks. The lower the entropy of the price return time series, the higher the price predictability. In particular, we examine the GameStop case, whose price increased significantly in January 2021. 

\subsubsection*{The GameStop case}
In early 2021, the eyes of everyone were on the New York Stock Exchange  because of one of the most terrific surges in prices in its history. Starting from the very end of December 2020, GameStop and multiple other stocks belonging to the same sector experienced a dramatic increase in their share price and volume, driven by the long trades of a huge number of individual investors, who fomented such a coordinated action on the Reddit social platform. When the prices were hitting their all-time highs, the attention of everyone was focused on such shares, which became known as "meme stocks".  Then, as the end of January 2021 approached, several retail broker-dealers temporarily limited certain operations in some of these stocks and options. As a consequence, the trend started to revert, even if such a turmoil period has had a permanent impact on GameStop and the other meme stocks. The stocks have maintained a higher price level with respect to the period before and have been characterized by higher volatility. A precise description of the GameStop case can be found in the report by the staff of the U.S. Securities and Exchange Commission \cite{sec_report}. There are two crucial aspects that have led to the sharp increase in meme stock prices. First, the aggregation of the orders sent to broker-dealers via online trading platforms, such as Robinhood, in the hands of a few off-exchange market makers permitted to negotiate good agreements, which have then resulted in incentives or no fee at all for the end customers.  Such an absence of trading frictions driven by FinTech innovations has had a positive feedback effect on retail trading. The second and more important aspect is the coordination in long trades by a huge number of individual investors, which has been possible because of online social networks like Reddit. The coordination has shaped up to be an act of rebellion against short-selling professional investors who had allegedly targeted the meme stocks. The two effects combined have created a first sharp increase in the prices, which has been further amplified by the coverage of the short positions by professional investors. The increasing interest in buying meme stocks and the resulting feedback dynamics have then led the intermediaries to extraordinary operations. Without entering into the discussion about the manipulative nature of such trading behavior coordinated via social networks and the interpretation of the GameStop rally as a revenge against Wall Street for the 2007-2008 crisis, it is evident that the case of meme stocks has represented a period when the market was out of the ordinary. Here, the focus is on market (in)efficiency during such  a period. In particular, we find that the clear predictability of the price pattern results in a signal of inefficiency in terms of the Shannon entropy and statistically significant variations of the Shannon entropy have an interpretation as early-warnings.

\subsubsection*{Structure of the paper}
Section \ref{Materials and Methods} presents the test of equality of entropies and a method for choosing the length of an interval for comparison. In particular, we give an estimation for the variance in Section \ref{Variance of the entropy's estimator}. We propose a method for choosing the length of a sliding window in Section \ref{Determining bandwidth}. We test the hypothesis about equal entropies on simulated data in Section \ref{Simulations}. The hypothesis is tested for the price returns of meme stocks in Section \ref{Entropy of price return time series}. The Appendix Sections A, B, and C contain the proofs of propositions and theorems. Section \ref{Conclusions} concludes the paper.

\section{Methodology and Dataset}
\label{Materials and Methods}
First, we discuss the Empirical Frequencies method for estimating the Shannon entropy. Then, we find the approximation of the variance of this estimation. We use the variance to make a hypothesis testing for equality of entropy values.
\subsection{Shannon entropy}
The Shannon entropy is defined as the average amount of information that a process transmits with each symbol \cite{Shannon}.
\begin{definition}
Let $X = \lbrace X_1 , X_2 , \dots\rbrace$ be a stationary random process with a finite alphabet $A$ and a measure $p$. A $k$-th order entropy of $X$ is
\begin{equation*}
\label{eq:entropy}
H_k(p)=-\sum_{x_1^k \in A^k}p(x_1^k)\log{p(x_1^k)}
\end{equation*}
where $x_1^k$ are all sequences of length $k$ with the convention $0\log{0}=0$. A process entropy of $X$ is
\begin{equation*}
h(p)=\lim_{k\to \infty} \frac{H_k(p)}{k}.
\end{equation*}
\end{definition}

We use the Empirical Frequencies method \cite{Marton} to estimate $k$-th order entropy from the sequence $x_1^n$. For each $a_1^k \in A^k$ empirical frequencies are defined as

$$
f(a_1^k|x_1^n)=\#\{i \in [1, n-k+1]: x_i^{i+k-1}=a_1^k \},
$$

where $x_i^{i+k-1}=x_i\dots x_{i+k-1}$. The estimation of the $k$-th order entropy is
\begin{equation}
\label{Eq: order entropy}
\hat{H}_k(x_1^n)=-\sum_{a_1^k}\hat{p}_k(a_1^k|x_1^n)\log{(\hat{p}_k(a_1^k|x_1^n))}
\end{equation}

where
\begin{equation*}
\label{empirical prob}
\hat{p}_k(a_1^k|x_1^n)=\frac{f(a_1^k|x_1^n)}{n-k+1}
\end{equation*}
The process entropy can be estimated as $\frac{\hat{H}_k}{k}$.
\subsection{Variance of the entropy's estimator}
\label{Variance of the entropy's estimator}

Let's assume that there are $M$ \textit{events} which can appear with probabilities $p_0, p_1, p_{M-1}$, $\sum_{j=0}^{M-1}p_j=1$. We assume that all $p_j$ are positive, because zero probabilities do not affect the entropy value $H=-\sum_{j=0}^{M-1}p_j\ln{p_j}$. If events appear independently $n$ times in total, the frequencies of events $f_0, f_1, f_{M-1}$ follow a multinomial distribution. Each frequency is distributed as Binomial $B(p_j,n)$. Therefore, the estimation of $p_j$, $\hat{p}_j=\frac{f_j}{n}$, is distributed as $B(p_j,n)/n$. The aim of this section is to find the variance of a random variable $\hat{H}=-\sum_{j=0}^{M-1}\hat{p}_j\ln{\hat{p}_j}$, where $\ln$ is the natural logarithm. The variable $\hat{H}$ is a $k$-th order entropy from Eq.~\ref{Eq: order entropy} with the base $e$ of the logarithm assuming that blocks $a_1^k\in A^k$ are generated independently. If all blocks have a non-zero probability, $M=|A|^{k}$, where $|A|$ is the size of the alphabet $A$.
\begin{theorem}[Approximation of variance]
Let's assume that $f_j$, $j=0\dots M-1$, are distributed as multinomial variables $f^M(p_0,\dots,p_{M-1},n)$ and $\hat{H}=-\sum_{j=0}^{M-1}\frac{f_j}{n}\ln{\frac{f_j}{n}}$. Then,
\begin{equation}
\label{eq:Theorem1}
Var(\hat{H})=\frac{1}{n}\left[-H^2+\sum_j p_j\ln^2(p_j)\right]+\frac{1}{n^2}\left[\frac{M}{2}-\frac{1}{2}\right]+\frac{1}{6n^3}\left[(1-H)\sum_j\frac{1}{p_j}-\sum_j{\frac{\ln{p_j}}{p_j}}-1\right]+O(n^{-4})
\end{equation}
\end{theorem}

where $H=-\sum_{j=0}^{M-1}p_j\ln{p_j}$. The proof of the Theorem is in Appendix \ref{Proof of the Theorem 1}. All useful propositions for proving the Theorem are in Appendix \ref{List of propositions}. In particular, we present a recursive formula for the central moments of the multinomial distribution of $(\hat{p}_1,\hat{p}_2)$ in Proposition \ref{proposition:1}. Also, we find the expression for the expectation of $\hat{p}^2\ln^2(\hat{p})$ that appears in $\hat{H}^2$ using the Taylor expansion in Proposition \ref{proposition:4}. Using all needed central moments of binomial and multinomial distributions, we calculate $Var(\hat{H})=E(\hat{H}^2)-E(\hat{H})^2$.

\begin{theorem}[Estimation of variance]
In addition to the assumption of Theorem 1, let's assume that all events appear at least once. Then, $Var(\hat{H})=E(\hat{Var})+O(n^{-4})$, where
\begin{equation}
\label{eq:Theorem2}
    \begin{split}
\hat{Var}&=\frac{1}{n}(\sum_j \hat{p}_j\ln^2{\hat{p}_j}-\hat{H}^2)+\frac{1}{n^2}\left[\sum_j \hat{p}_j\ln^2{\hat{p}_j}-\hat{H}^2-M\hat{H}-\sum_j{\ln{\hat{p}}}-\frac{M}{2}+\frac{1}{2}\right]\\
&+\frac{1}{n^3}\left[\sum_j \hat{p}_j\ln^2{\hat{p}_j}-\hat{H}^2-M\hat{H}-\sum_j{\ln{\hat{p}_j}}-\frac{\hat{H}}{3}\sum_j{\frac{1}{\hat{p}_j}}-\frac{1}{3}\sum_j{\frac{\ln{\hat{p}_j}}{\hat{p}_j}}-\frac{1}{12}\sum_j{\frac{1}{\hat{p}_j}}-\frac{M^2}{4}-\frac{M}{2}+\frac{5}{6}\right]
    \end{split}
\end{equation}
\end{theorem}

The proof of Theorem 2 is in Appendix \ref{Proof of Theorem 2}. In the Theorem, we assume that all different events appear in the sequence, thus $M$ is known as the number of all different events. We discuss this assumption in the Remark below.

\begin{remark}
Number of appearing different events, $\hat{M}$, is defined as $\hat{M}=M-\sum_{j=0}^{M-1} I\{\hat{p}_j=0\}$, where $I$ is an indicator function. Thus,

$$E[\hat{M}]=M-\sum_{j=0}^{M-1} (1-{p_j)^{n}}.$$
\end{remark}

The error term $\sum_{j=0}^{M-1} (1-{p_j)^{n}}$ attains its minimum when all $p_j=\frac{1}{M}$. It grows as a probability approaches 0; that is, when there is an event that can happen with a tiny but not zero probability, so that we may not observe it in a finite sequence of events. Such an event does not greatly affect the entropy value, since $p\ln{p}\to 0$ as $p\to 0$. We need to fix the minimum length of a sequence to assume that all events appear in the sequence. To this aim, we introduce the following rule. The length of the sequence is taken such that the minimum error is less than 0.01. That is, one event with a non-zero probability does not appear in 1 case out of 100.

\begin{equation*}
    \begin{split}
        \sum_{j=0}^{M-1} (1-{p_j)^{n}}<0.01\\
        M(1-{\frac{1}{M})^{n}}<0.01\\
        n>\frac{\ln{\frac{0.01}{M}}}{\ln{\frac{M-1}{M}}}\\
        n_{min}=\lceil{\frac{\ln{\frac{0.01}{M}}}{\ln{\frac{M-1}{M}}}}\rceil
    \end{split}
\end{equation*}
\subsection{Hypothesis testing}
Using the estimation of entropy, we aim to conclude if two entropy values significantly differ. Let's take two sequences with entropies $H_1$ and $H_2$. Let $\hat{H}_1$ and $\hat{H}_2$ be estimations of entropies with variances $Var_1$ and $Var_2$, respectively.
\begin{equation*}
    \begin{split}
\mathcal{H}_0: H_1=H_2\\
\mathcal{H}_a: H_1\neq H_2
    \end{split}
    \end{equation*}

Let assume that $H_1=H_2$. Then, a z-score

\begin{equation}
\label{eq:z}
z=\frac{\hat{H}_2-\hat{H}_1}{\sqrt{Var_2+Var_1}}
\end{equation}

is distributed with a zero mean and the variance equal to 1. We reject $\mathcal{H}_0$ if $|z|$ is larger than a quantile corresponding to $99\%$ of confidence. The quantile is defined empirically in Section \ref{Empirical quantiles}.
\subsection{Determining bandwidth}
\label{Determining bandwidth}
We assume that a time series into consideration can be not stationary, thus we need to choose a \textit{bandwidth} $w$, that is the length of a rolling window where entropy is estimated. We aim to detect significant changes in entropy, thus we can not take $w$ too large so that the window covers several intervals with different entropy values. On the other hand, if the process at some period is stationary, we aim to take $w$ as large as possible to improve the accuracy of the entropy estimation\footnote{The equations (\ref{eq:Theorem1}) and (\ref{eq:Entropy}) show that the larger the length of a sequence, the less the variance and the downward bias of the entropy estimation.}. In terms of the bias-variance trade-off, if the rolling window covers a period where the process is stationary (with constant entropy), the error from a bias is eliminated. However, taking $w$ too small implies a large variance and thus it may be impossible to distinguish a change in entropy from estimation errors. Thus, our goal is to find an optimal parameter $w$.

We select the following approach to determine the optimal bandwidth. We test for all adjacent non-overlapping intervals. Then, we choose a bandwidth that allows us to detect the maximum z-score. More precisely, we aim to maximize an objective function $f(w)$ below.\footnote{To find the maximum, we use the function scipy.optimize.minimize\_scalar with method=bounded and xatol= 1 in Python v.3.9.5.}

\begin{equation}
\label{eq:f}
    f(w)=\begin{cases}
    \max(|z(w)|)\text{, if }\%\{|z(w)|>q_{99}\}>1\%\\
    -\frac{1}{w}\text{, otherwise}
    \end{cases}
\end{equation}

where $q_{99}$ is a $99\%$ quantile for the empirical distribution of $z$. The intuition for the objective function is that small values of $w$ give small values of the z-score since the variance, $Var(\hat{H})$, is $O(w^{-1})$. Large values of $w$ may not be able to catch the largest change in entropy value. Finally, if we can not reject the hypothesis about constant entropy, we aim to choose $w$ as large as possible, thus we maximize $-\frac{1}{w}$. Since we are interested in detecting changes in entropy, $\max(|z|)$ is always non-negative and $-\frac{1}{w}$ is always negative. We give more intuition in Section~\ref{Testing of a non-stationary process} by plotting Fig.~\ref{fig:paper3_inefflengths}.

To make at least one test, we set the upper bound for the bandwidth as $n_{max}=\lfloor\frac{n}{2}\rfloor$, where $n$ is the total length of the sequence. We may define the optimal bandwidth using a training set.
\subsection{Dataset}
We consider three meme stocks that became popular in late 2020 and 2021. We take a one-minute frequency from 9:00 to 15:59. We investigate stocks of the companies GameStop (GME), Bed Bath \& Beyond (BBBY), and AMC Entertainment Holdings (AMC)\footnote{We use a proprietary intraday financial time series dataset provided by kibot.com.}. In addition, we consider three well-known IT companies, expecting more persistent entropy for their stocks. These companies are Apple (AAPL), Salesforce (CRM), and Microsoft (MSFT). Using return time series, we define a 4-symbols alphabet as follows.

\begin{equation}
\label{discretization}
\begin{split}
s_t=
\begin{cases}
0, r_t\le Q_1, \\
1, Q_1<r_t\le Q_2,\\
2, Q_2<r_t\le Q_3,\\
3, Q_3<r_t,
\end{cases}
\end{split}
\end{equation}

where $Q_1$, $Q_2$, $Q_3$ are quartiles of the empirical distribution of returns $r_t$. Here, $r_t$ are price returns time series after filtering out data regularities: intraday volatility pattern, heteroskedasticity, price staleness, and microstructure noise \cite{Shternshis2}. Price returns are defined as $R_t=\ln{\frac{P_t}{P_{t-1}}}$, where $P_t$ is a price at time $t$. Data regularities are empirical properties of price returns \cite{Cont} that make the price returns time series more predictable but do not imply any profitable trading strategy. For instance, intraday volatility pattern refers to the fact that the volatility of intraday returns has periodic behavior. It is higher near the opening and the closing of the market \cite{Wood}, because traders tend to trade less in the middle of a day.

\section{Simulation study}
For the further analysis of simulated and real data, we fix $A=\{0,1,2,3\}$.
\label{Simulations}
\subsection{Empirical quantile}
\subsubsection{Determining quantile of entropy's distribution}
\label{Empirical quantiles}

In this section, we find the quantiles of the distribution of the z-score from Eq.~\ref{eq:z}. Intuitively, the larger $M$ (the number of different events), the more the entropy estimator (as the sum of random variables) tends to be normally distributed. G.~P.~Basharin showed that the distribution of the entropy estimator is asymptotically normal \cite{basharin59}. However, if all probabilities are almost equal, the distribution of scaled entropy estimation converges to $\chi^2$-distribution \cite{Zubkov74}. That is, quantiles of normal distribution can be used if the entropy is not near the maximum. In case when entropy is close to the possible maximum, the normal distribution for testing equality of two entropies is inappropriate.\footnote{The difference between two $\chi^2$-distributions we are interested in is no longer a $\chi^2$-distribution. The density function of the difference takes a complex form and is derived in the article \cite{Mathai93} (Theorem 2.1). Making a statistical test based on $\chi^2$-distribution of entropy estimation without subtraction is postponed for future research.} To show this, we set large values for the length of a sequence and the length of blocks.

We perform $N=2\times10^4$ Monte Carlo simulations. We simulate two sequences with length $n=2\times10^5$ of 4 symbols with equal probabilities. We set $k=7$ so that $M=4^7=16384$. When probabilities are equal, the expression for the variance (Eq.~\ref{eq:Theorem1}) becomes

$$Var(\hat{H}_{max})=\frac{M-1}{2(n-k+1)^2}+\frac{M^2-1}{6(n-k+1)^3}.$$

We use this value as given instead of estimating the variance from the sequence. Then, for two sequences we find the value of $\Delta \hat{H}=\hat{H_2}-\hat{H_1}$. Since $\Delta \hat{H}$ is the difference of two independent variables, its variance is $2Var(\hat{H})$. An empirical $99\%$ ($95\%$) quantile of the empirical distribution of  $|z|$ is $3.30722$ ($2.54542$). Thus, even if the length of a sequence and the amount of blocks are quite large, the tails of the empirical distribution are thicker than for the normal distribution. We use the empirical quantiles for further analysis. If an absolute value of z-score is larger than the quantile, the difference between entropy values is statistically significant.
\subsubsection{Testing short independent sequences}
For the further analysis, we fix $k=4$. Therefore, the maximum of the $k$-th order entropy is $k\ln{|A|}=4\ln{4}$, $M=256$.

Now, we aim to test the quantiles found in the previous section for shorter sequences. Here, we keep $N=2\times10^4$ and set $n=2\times10^3$. Using quantiles from the previous section, the false positive rate is $1.025\%$ ($4.86\%$) for the level of significance $\alpha=0.01$ ($\alpha=0.05$).\footnote{Empirical quantiles from this experiment are equal to $3.31684$ and $2.52907$, respectively.} We consider these results as acceptable for keeping the found values of quantiles for the rest of the paper.
\subsubsection{Testing a fully random sequence}
Here, we simulate one sequence with four equiprobable symbols and length $N=2\times10^7$ and divide it by overlapping intervals with the length $n=2\times10^3$. We consider all differences with the gap equal to $n$, so that there are no common blocks between two intervals. The false positive rate according to the $99\%$ ($95\%$) empirical quantile is $0.9976\%$ ($4.4608\%$), that is quite close to the significance level. Thus, $q_{99}$ from Eq.~\ref{eq:f} is taken as $3.30722$.
\subsection{Power and size of the test}
We consider a process with 4 symbols. The probability of repeating a symbol is $\tau$. All other probabilities are equal, that is, the probability of having, for example, 0 after 1 is $\frac{1-\tau}{3}$. Thus, the value of the 4th order entropy is given by 

\begin{equation}
\label{eq:Htau}
    H(\tau)=-2\left(\tau\ln{\frac{\tau}{4}}+(1-\tau)\ln{\frac{1-\tau}{12}}\right).
\end{equation}
    
The further $\tau$ from $\frac{1}{4}$, the lower the entropy. $\tau=\frac{1}{4}$ corresponds to equiprobable symbols and the maximum of entropy. First, we construct two sequences with $H(\frac{1}{4})$ and length $n=10000$. We estimate the variance of entropy's estimator using Equation \ref{eq:Theorem2} and perform hypothesis testing using Equation \ref{eq:z}. The false positive rate (size) is $0.86\%$ obtained by generating two sequences $2\times10^{4}$ times. Now, we simulate one sequence with $H(\frac{1}{4})\approx5.54518$  and the other with a different entropy value to measure the power of the test. The results are in Table \ref{Table: power}.
\begin{table}[ht]
\caption{The power of the hypothesis testing with different $\tau$.}
\centering
\begin{tabular}{|l|l|l|}
\hline
 $\tau$&$H(\tau)$  &Power (\%)  \\ \hline
 0.28&5.5405  &56.28  \\ \hline
 0.29&5.53692  &94.556  \\ \hline
 0.3&5.53237  &99.915  \\ \hline
 0.31&5.52688  &100  \\ \hline
\end{tabular}
\caption*{Power is the probability to reject $\mathcal{H}_0$ given that the null hypothesis is false. The results are obtained with $2\times10^{4}$ Monte Carlo simulations.}
\label{Table: power}
\end{table}
\subsection{Non-stationary process}
\label{Testing of a non-stationary process}
\begin{figure}[ht]
\centering
\includegraphics[width=8 cm]{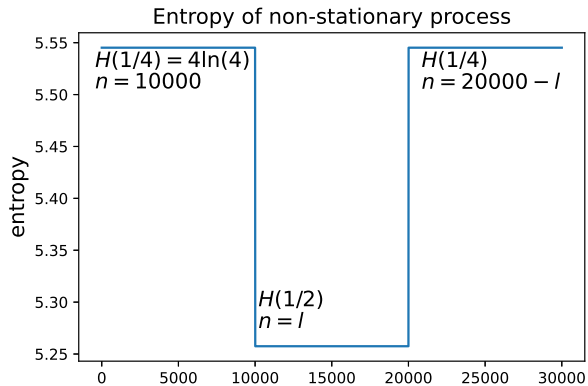}
\caption{Illustration of stepwise entropy.\label{fig:nostationaryentropy}}
\end{figure}

Now, we consider a process with the 4-symbols alphabet constructed as follows. The length of a sequence is $N=30000$. We set $k=4$, $M=256$. For this sequence, $n_{min}=\lceil{\ln{\frac{0.01}{M}}}/{\ln{\frac{M-1}{M}}}\rceil=2594$ and $n_{max}=\lfloor\frac{N-k+1}{2}\rfloor=14998$. We divide the sequence into three parts as shown in Fig.~\ref{fig:nostationaryentropy}. The first and the last parts are simulated as the processes with the maximum entropy and four equiprobable symbols. The first part has the length of $10000$. The part with the length $l$ in the middle has the entropy $H(\tau)$ given by Eq.~\ref{eq:Htau}.

\begin{figure}[ht]
\centering
\includegraphics[width=9 cm]{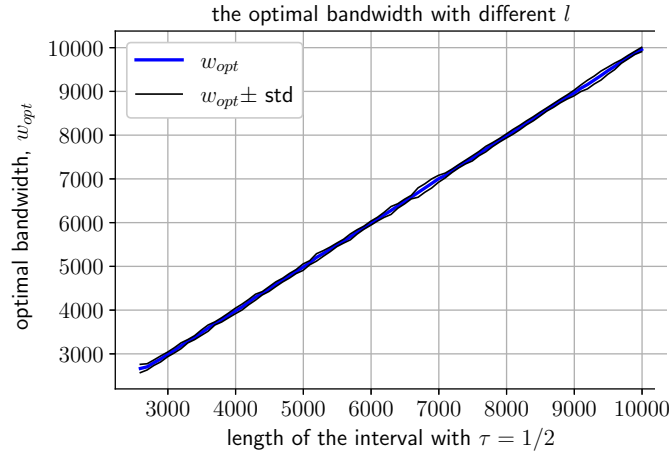}
\caption{Optimal bandwidth for different values of $l$. The mean and standard deviation (std) are calculated over 100 iterations.\label{fig:bandwidth_l}}
\end{figure}

\begin{figure}[htb]
\centering
\includegraphics[width=7 cm]{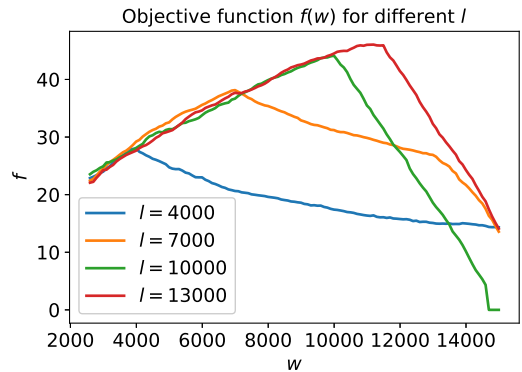}
\caption{Objective function for different values of $l$.\label{fig:paper3_inefflengths}}
\end{figure}

\begin{figure}[htb]
\centering
\includegraphics[width=7 cm]{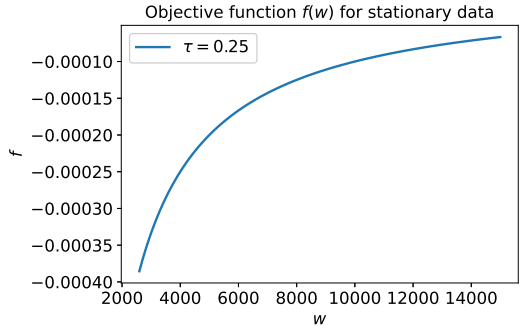}
\caption{Objective function for one realization of stationary process.\label{fig:paper3_kernel_efflength3}}
\end{figure}

The variance of the entropy's estimator becomes time-varying, thus we estimate it using Equation \ref{eq:Theorem2}. We apply the method for determining an \textit{optimal bandwidth}, $w_{opt}$, from Section~\ref{Determining bandwidth}. First, we  fix $\tau=0.5$ and take $l$ from $n_{min}$ to $10000$ with the increment of $100$. For each $l$, we calculate the optimal bandwidth and plot it in Fig.~\ref{fig:bandwidth_l}. The figure shows that the optimal bandwidth corresponds to the length of the interval in the middle. Fig.~\ref{fig:paper3_inefflengths} shows the plot of the objective function (Eq.~\ref{eq:f}) for 4 random iterations with different values of $l$. All plots have one global maximum. The larger $l$, the larger the argument of the maxima corresponding to $w_{opt}$. When $l=10000$ and $w$ is close to $n_{max}$, the percent of statistically significant changes in entropy is less than 1\%, thus $f(w)$ becomes negative. If the process generating the sequence is stationary ($\tau=0.25$), the range of the objective function may be negative. The objective function of one realization of the stationary process is given in Fig.~\ref{fig:paper3_kernel_efflength3}. Since the function is monotonically increasing, the maximum is attained at $n_{max}$. 

\begin{figure}[htb]
\centering
\includegraphics[width=7 cm]{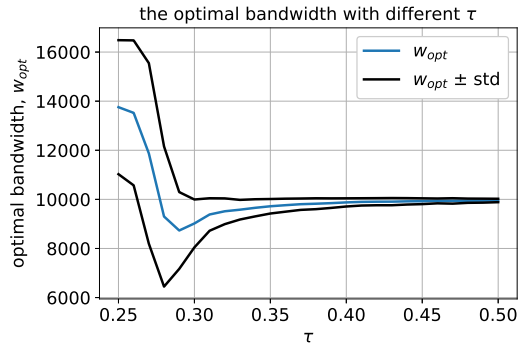}
\caption{Optimal bandwidth for different values of $\tau$. The mean and standard deviation (std) are calculated over 400 iterations.\label{fig:bandwidth_alphas}}
\end{figure}

Finally, we fix $l=10000$ and take $\tau$ from $0.25$ to $0.5$ with the increment of $0.01$. For each $\tau$, we calculate the optimal bandwidth and plot it in Fig.~\ref{fig:bandwidth_alphas}. $\tau=0.25$ corresponds to the case of a stationary process and thus $w$ is close to the possible maximum. This corresponds to the intuition of choosing the bandwidth: if the process is stationary, the larger the bandwidth, the more accurate the entropy estimate. The deviation of $w_{opt}$ when $\tau=0.25$ is high because of the first type error in hypothesis testing. Even if the process is stationary, $\mathcal{H}_0$ may be rejected more than in $1\%$ of cases that leads to a positive value of $f(\omega)$. Such a large deviation can be reduced using Bonferroni or Šidák corrections \cite{Sidak67} if the distribution of a $z$-score is known. We note that when the entropy in the middle slightly differs from the maximum ($\tau=0.26$), the method may not detect the interval with different entropy and set $w_{opt}$ close to the maximum. The larger $\tau$, the closer $w_{opt}$ to the length of the interval in the middle and the less the standard deviation of $w_{opt}$.

\section{Empirical application: the case of meme stocks}
\label{Entropy of price return time series}
\subsection{Market efficiency}
A market in which prices always fully reflect available information is called efficient \cite{Fama}. In a weak form of the efficient market hypothesis, the information set is historical prices. Thus, if a market is efficient in the weak form, a future price can not be predicted better than its current value. Full uncertainty about future values of prices implies the maximum degree of randomness of the price returns time series. Thus, the entropy of return time series should attain its maximum if the market is efficient.

If a market is efficient and the entropy of the price returns is always at the maximum, the hypothesis $\mathcal{H}_0$ should be failed to reject. The rejection of $\mathcal{H}_0$ for two non-overlapping intervals implies that the entropy is time-varying. Low values of entropy relatively to entropy estimated in the past indicate predictability of price returns.
\subsection{Meme stocks}
Here, we consider stocks GME, BBBY, and AMC. We set the year 2019 as a training set. The period from 01.01.2020 to 20.07.2021 is a testing set.  First, we filter out the data regularities. We define an intraday volatility pattern, fit an autoregressive moving average (ARMA) model, and find an optimal bandwidth, $w_{opt}$, using the training set. Volatility and the degree of price staleness are defined minute by minute. The ARMA model is needed to filter out microstructure noise; volatility estimation is needed to filter out heteroskedasticity. Quartiles $Q_1$, $Q_2$, $Q_3$ used for discretization in Eq.~\ref{discretization} are defined using the return time series of the training set after filtering out the data regularities.

We calculate the Shannon entropy of the discretized sequence $s_t$ of the testing set using the rolling window with length $w_{opt}$. Comparing entropies of two adjacent intervals, we have three possible outputs: entropy decreases, entropy increases, entropy does not change significantly. We present results for each stock in Table~\ref{Table:1}.

We plot the entropy estimation, price, and trading volumes in Figure \ref{GME stock} for the stock GME. Plots of BBBY and AMC stocks are in Appendix \ref{Plots}. We mark on the plots where entropy has significant changes. Red dots correspond to statistically significant decrease in entropy. Green dots stands for statistically significant increase in entropy. The dots at the same time are plotted in figures for the prices and trading volumes. We report below when entropy started to decrease first time or after an increase. We note that the entropy had been already defined as statistically significant low for the stock AMC when record volumes occurred and the price rose sharply for the first time in January 2021. 
\begin{figure}[ht]%
\centering
\includegraphics[width=9 cm]{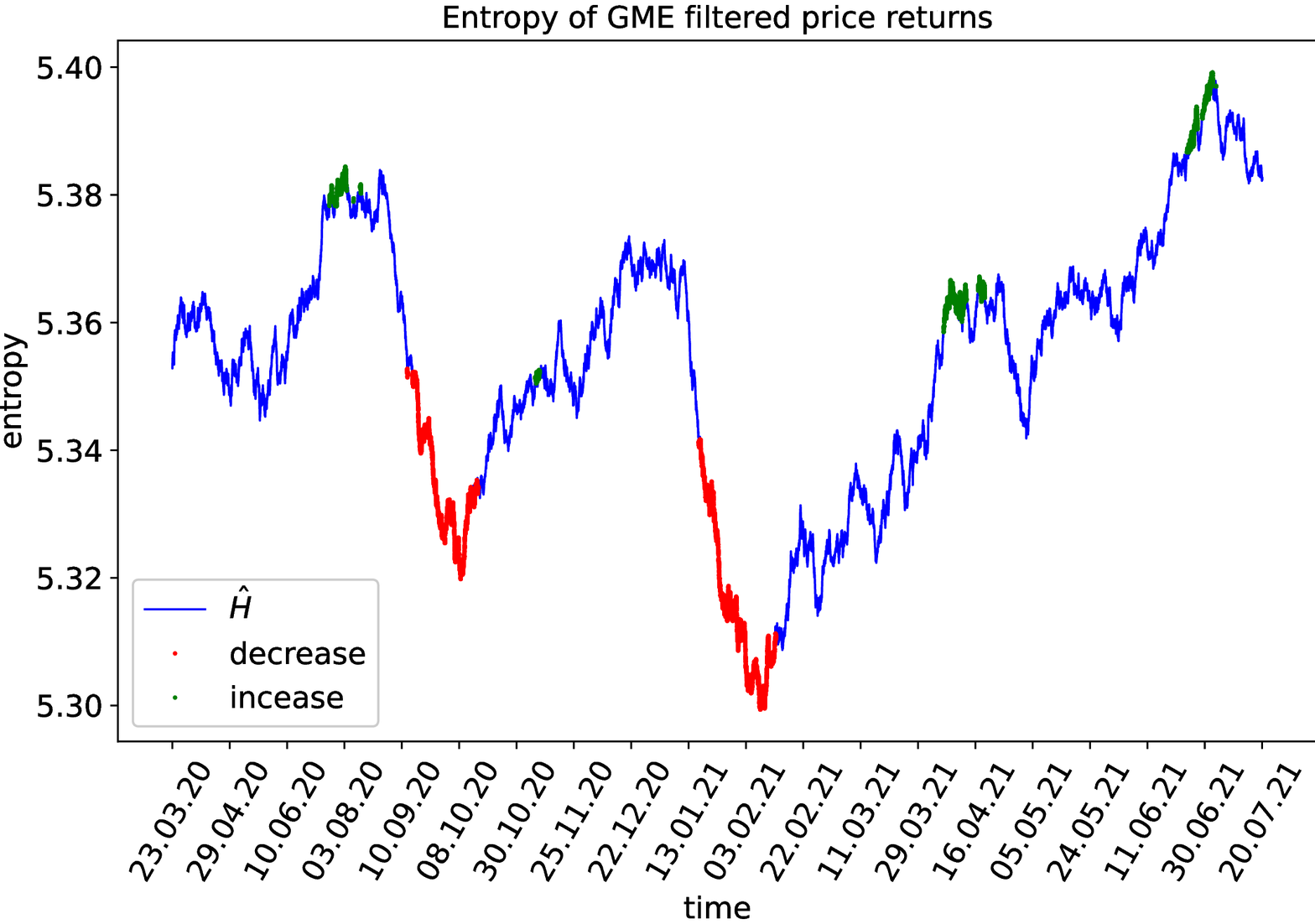}
\qquad
\includegraphics[width=9 cm]{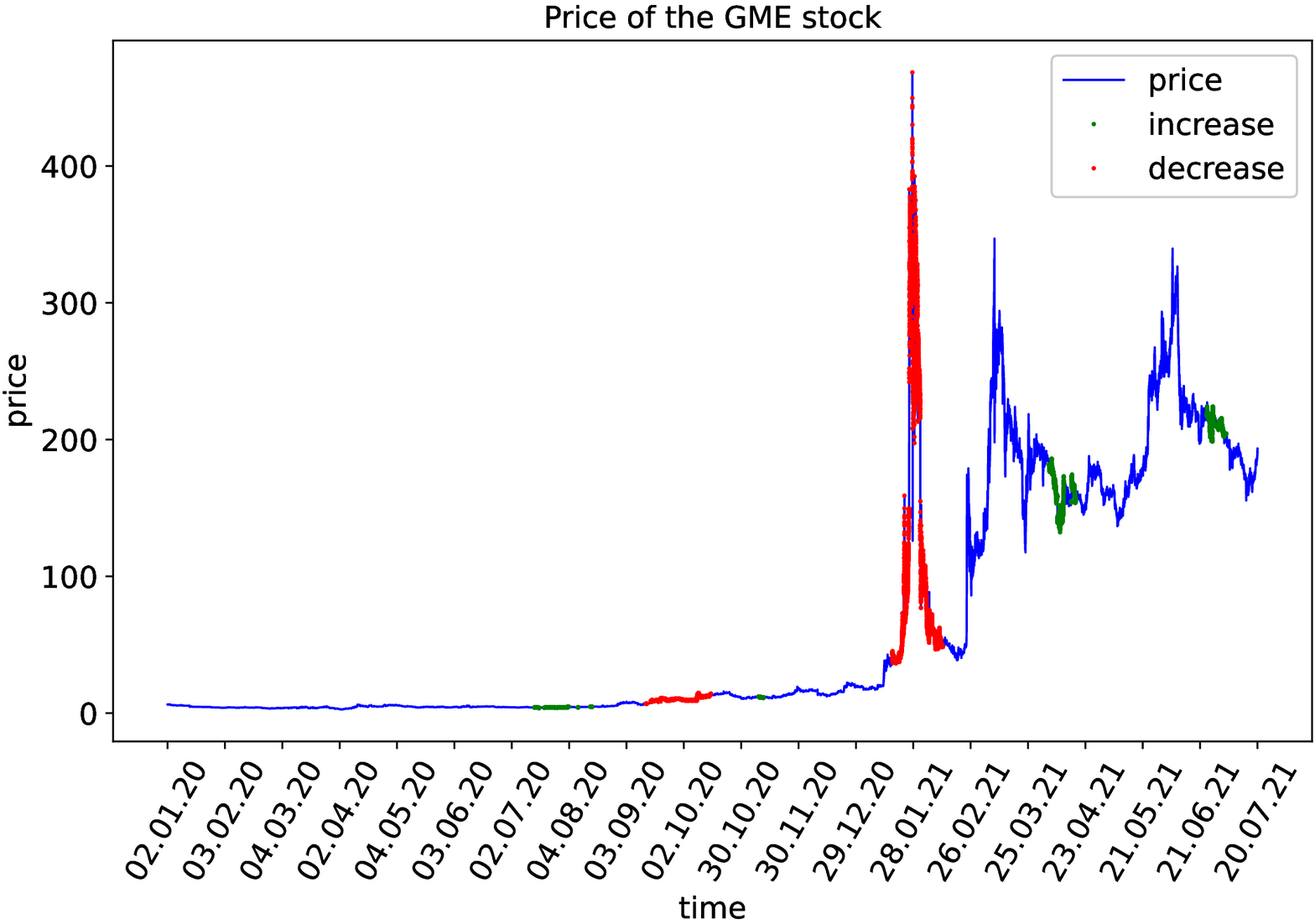}
\qquad
\includegraphics[width=9 cm]{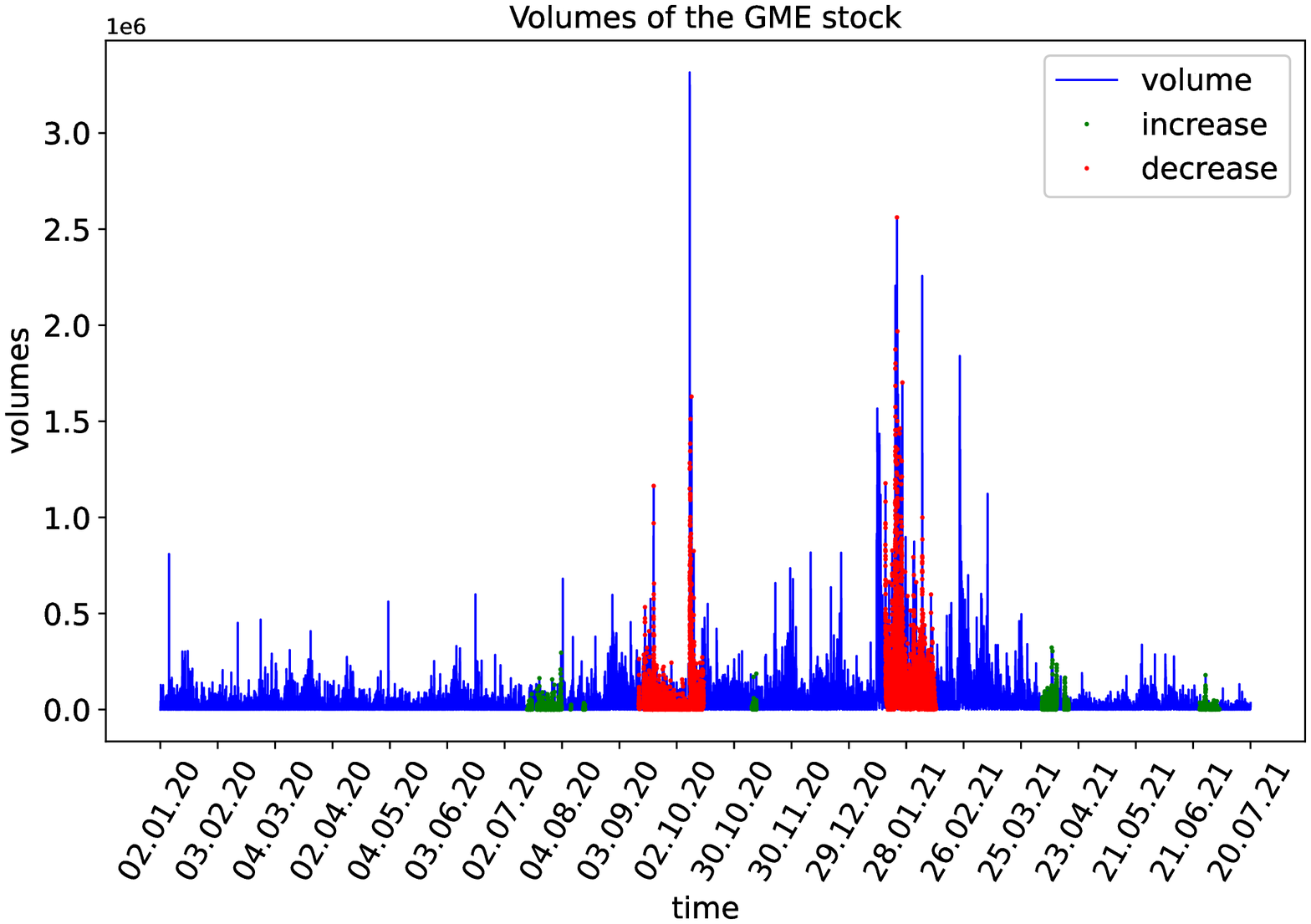}
\caption{Entropy, price, volume of the GME stock. Dots correspond to statistically significant changes in entropy.}
\label{GME stock}
\end{figure}
\clearpage

For GME, there are two series of decreases. They start from the entropies calculated at the periods [17.07.20 15:46 to 14.09.20 15:34] and \textbf{[7.12.20 15:38 to 19.01.21 09:42]}. For BBBY, there are three series of decreases. They start from the entropies calculated at the periods [06.03.20 13:50 to 05.05.20 09:54], \textbf{[10.12.20 11:47 to 28.01.21 10:59]}, and [06.05.21 15:16 to 21.06.21 14:45]. For AMC, there are four series of decreases. They start from the entropies calculated at the periods [05.03.20 15:13 to 17.04.20 12:47], [10.08.20 13:08 to 08.09.20 11:27], [08.03.21 14:09 to 30.03.21 10:26], and \textbf{[05.05.21 14:46 to 27.05.21 15:10]}. The time intervals highlighted in bold correspond to sharp increases in price values.

\subsection{IT stocks}
\begin{figure}[ht]%
    \centering
    \subfloat[AAPL\centering]{{\includegraphics[width=8 cm]{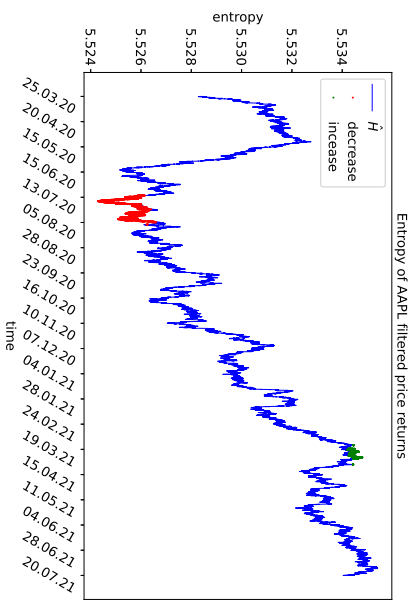} }}%
    \qquad
    \subfloat[CRM\centering]{{\includegraphics[width=8 cm]{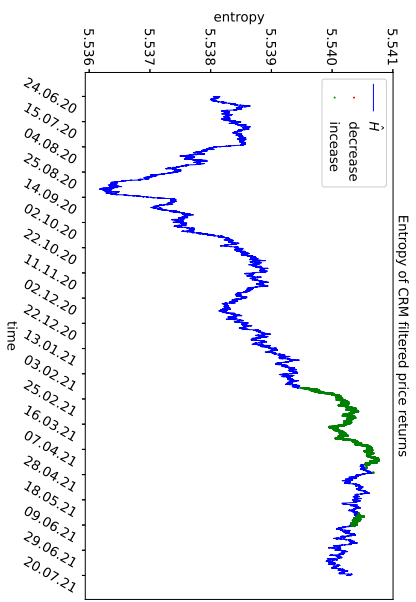} }}%
    \qquad
    \subfloat[MSFT\centering ]{{\includegraphics[width=8 cm]{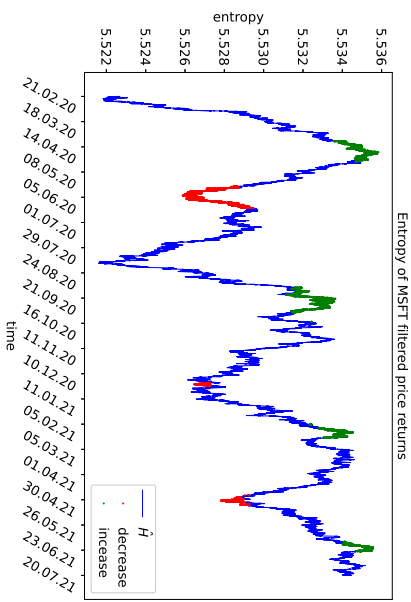} }}%
    \caption{Entropy of IT Stocks. Dots correspond to statistically significant changes in entropy.}%
    \label{fig: Entropy of IT Stocks}
\end{figure}

We take the stocks of Apple, Salesforce, and Microsoft for the comparison. Calculated entropies are plotted in Fig.~\ref{fig: Entropy of IT Stocks}; the optimal bandwidths are recorded in Table~\ref{Table:1}. The entropy of the price returns of these stocks exhibits time-varying behavior. For each stock, statistically significant changes in entropy are detected. However, the entropy does not fall as much as in the case of meme stocks. The price returns time series for the stock CRM in 2019 is defined as stationary since $f(w_{opt})$ is negative.
\begin{table}[ht]
\caption{Optimal bandwidth and hypothesis testing for meme and IT stocks.}
\begin{tabular}{|p{1cm}|p{1cm}|p{2cm}|p{1cm}|p{2cm}|p{2cm}|p{2cm}|}
\hline
Stock & $w_{opt}$ & $f(w_{opt})$ & $n_{max}$ & Number of tests & Number of increases & Number of decreases \\ \hline
GME   &6707            &3.881               &14387            &89909                 &5672                     &12468                     \\ \hline
BBBY  & 10218      & 4.975         & 22542      & 89248           & 7399                & 7831                \\ \hline
AMC   & 5845       & 5.052         & 16374      & 93639           & 6372                & 8107                     \\ \hline
AAPL  &20313            &9.323               &33865            &104531                 &1495                     &6137                     \\ \hline
CRM  &44321            &-2.256$\cdot 10^{-5}$               &44322            &55189                 &17651                     & 0                    \\ \hline
MSFT  &12539            &5.438               &41413            &121887                 &15969                     &7271                     \\ \hline
\end{tabular}
\caption*{$f(w_{opt})$ is calculated on the training set. The number of tests is the amount of adjacent time intervals with lengths $w_{opt}$ in the testing set. The increases and decreases are statistically significant changes in entropy between two adjacent intervals.}
\label{Table:1}
\end{table}

\clearpage
\subsection{Quarterly training sets}
\begin{figure}[ht]%
    \centering
    \subfloat[GME\centering]{{\includegraphics[width=8 cm]{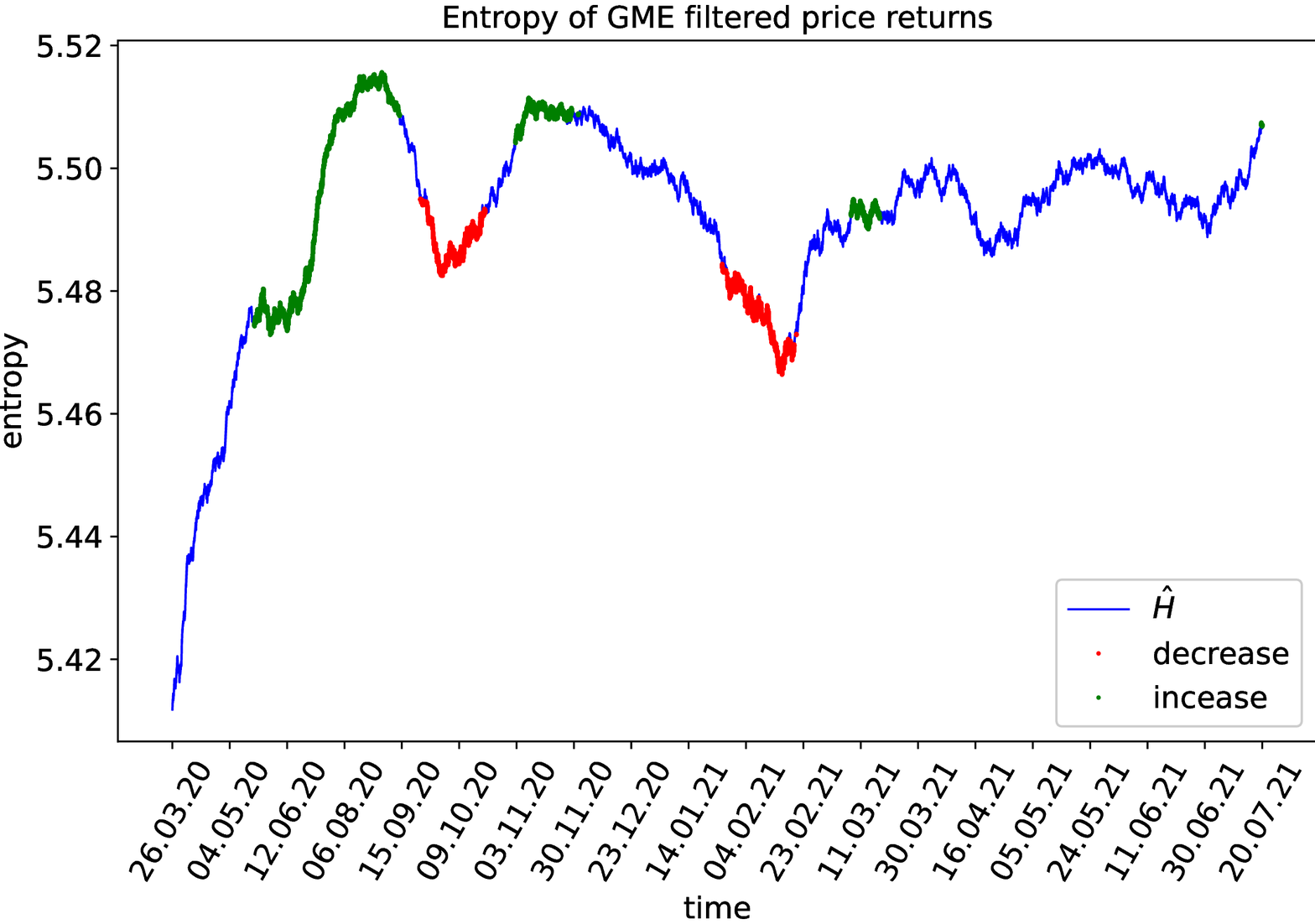} }}%
    \subfloat[BBBY\centering]{{\includegraphics[width=8 cm]{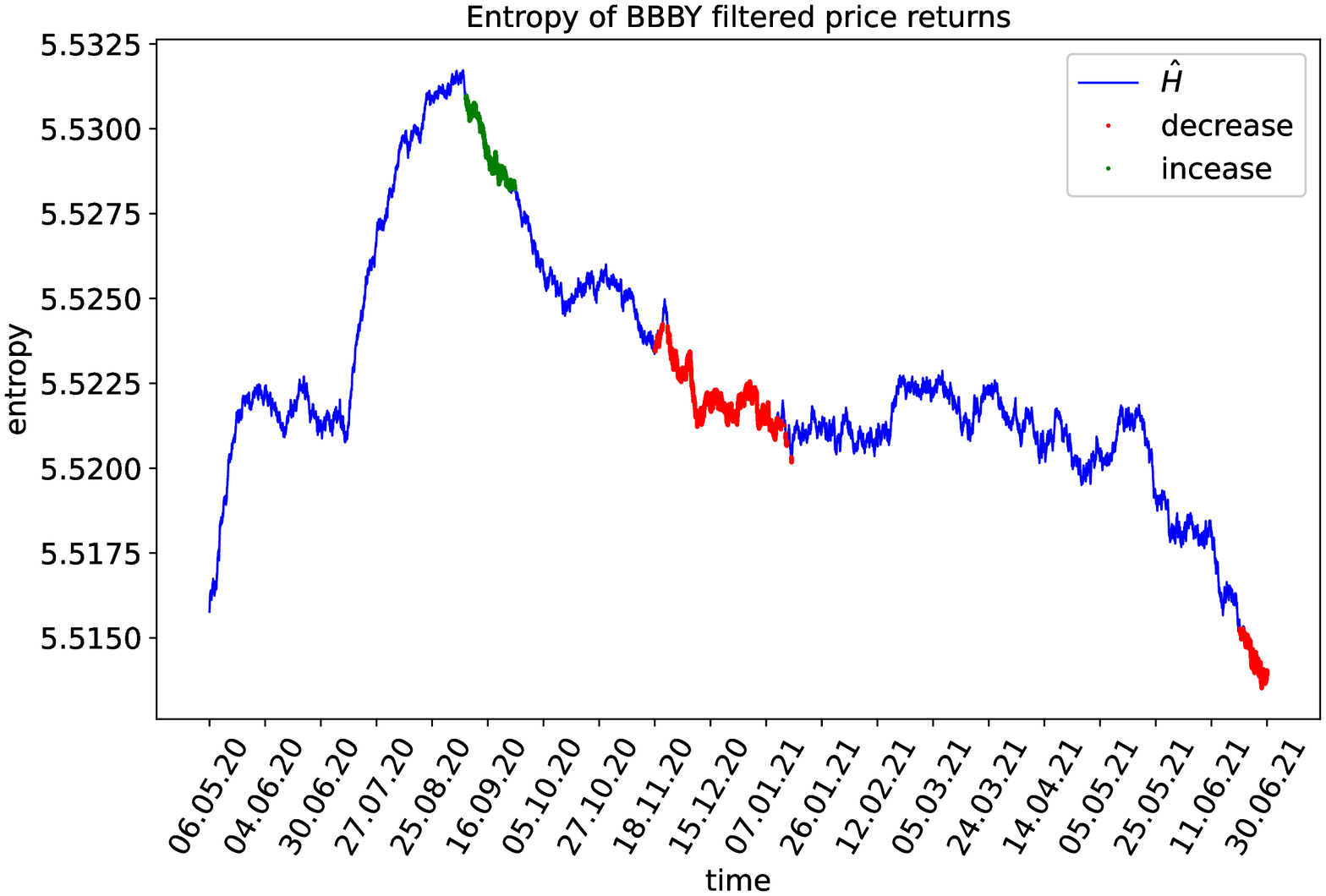} }}%
    \qquad
    \subfloat[AMC\centering ]{{\includegraphics[width=8 cm]{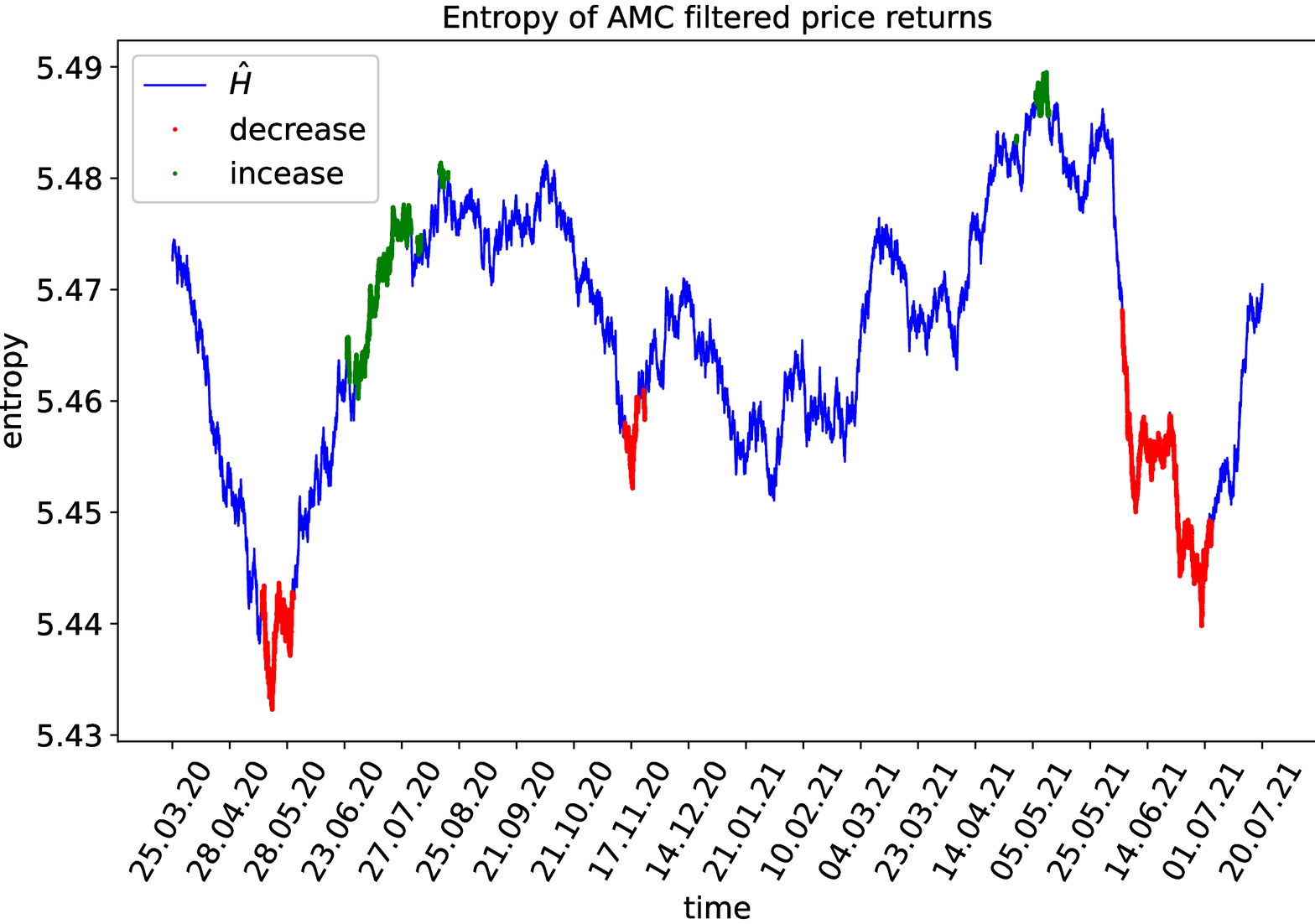} }}%
    \subfloat[AAPL\centering]{{\includegraphics[width=8 cm]{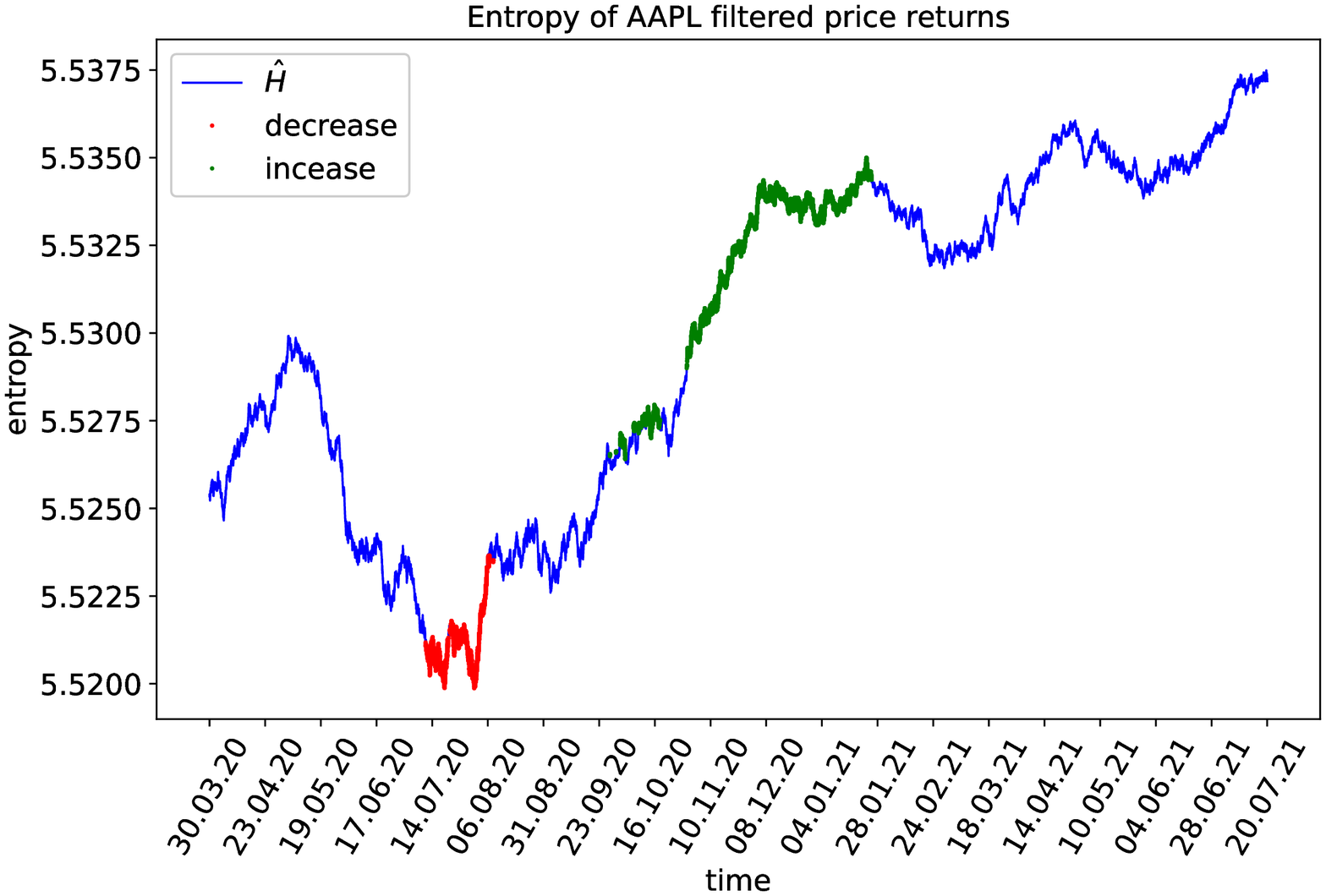} }}%
    \qquad
    \subfloat[CRM\centering]{{\includegraphics[width=8 cm]{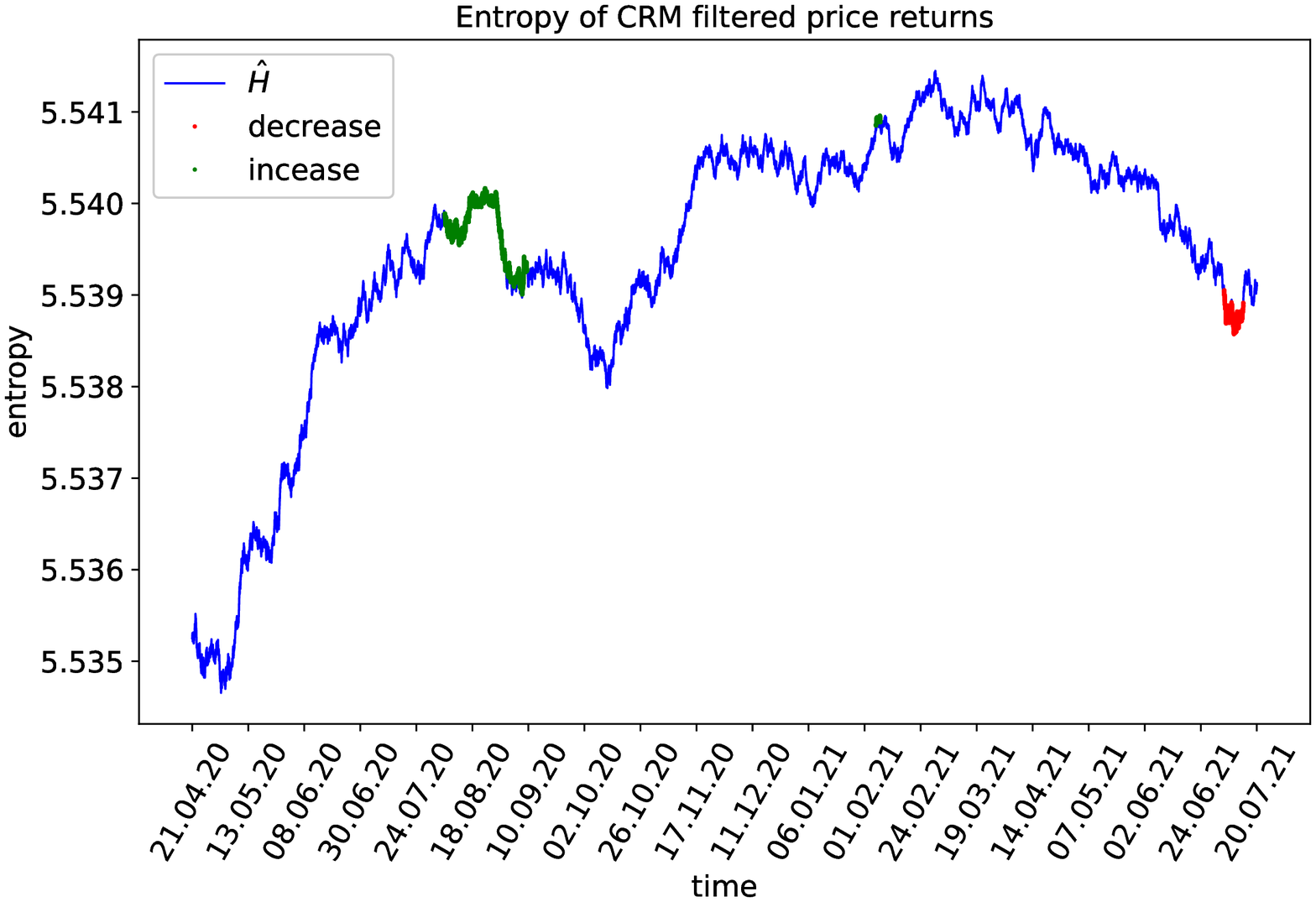} }}%
    \subfloat[MSFT\centering ]{{\includegraphics[width=8 cm]{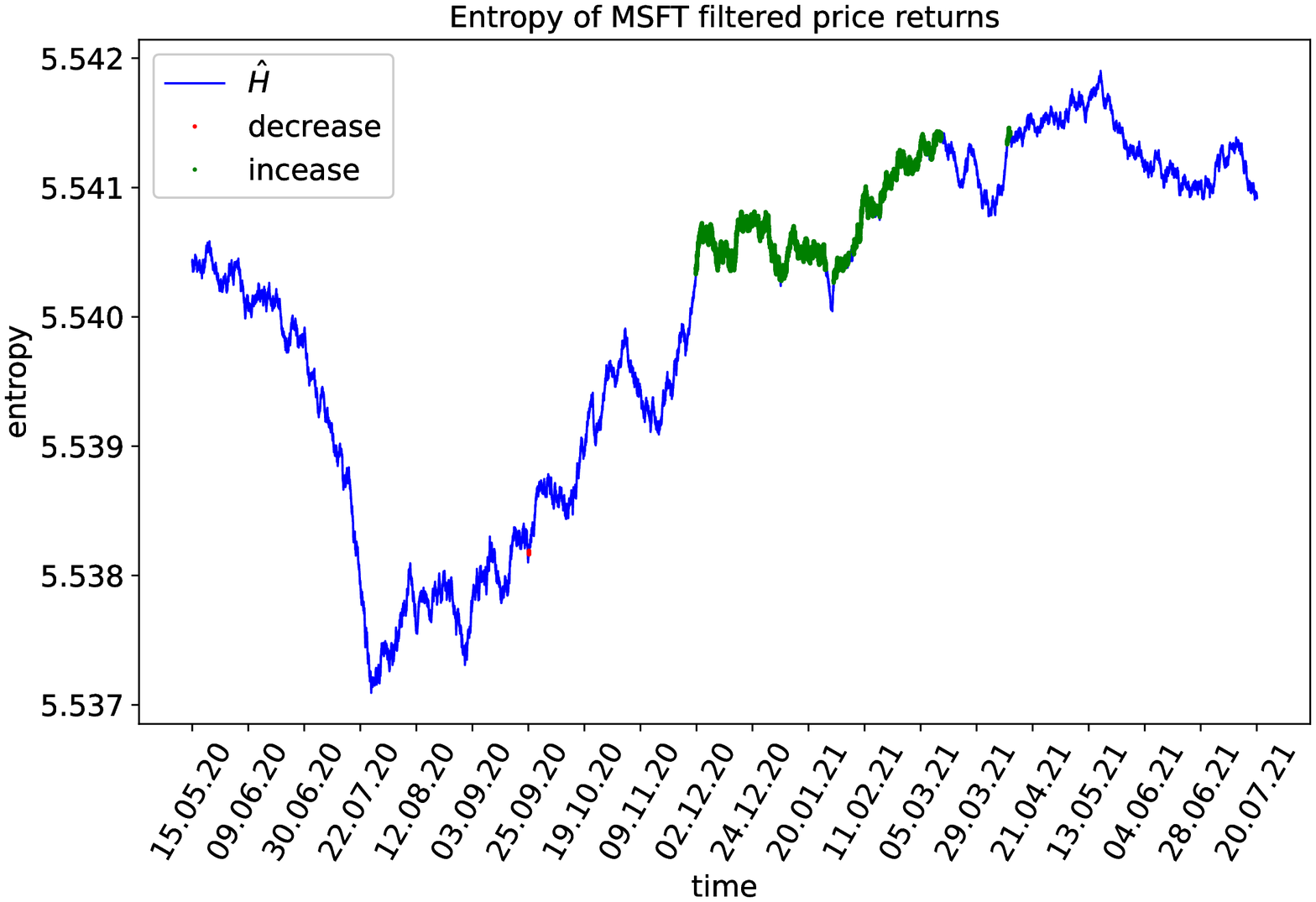} }}%
    \caption{Entropy of Stocks with quarterly training sets. Dots correspond to statistically significant changes in entropy.}%
    \label{fig: Entropy of Stocks with quarters}
\end{figure}

The reason for the increase in predictability, when entropy falls, may be a change in the structure of data regularities. For instance, due to the growing popularity of meme stocks, the intraday volatility pattern could change from the year 2019 to the year 2021. If the behavior of traders has changed, the intraday volatility pattern from the training set does not filter out the data regularity in the testing set. In order to filter the data regularities more carefully, we update an estimation of intraday volatility pattern and a fitted ARMA model using quarterly intervals starting from the first quarter of 2019. This quarter is also used to filter out the data regularities in-sample. An optimal bandwidth is defined using the year 2019. The results are shown in Table \ref{Table:2} and Fig.~\ref{fig: Entropy of Stocks with quarters}. 

\begin{table}[ht]
\caption{Optimal bandwidth and hypothesis testing for stocks using quarterly training sets}
\begin{tabular}{|p{1cm}|p{1cm}|p{2cm}|p{1cm}|p{2cm}|p{2cm}|p{2cm}|}
\hline
Stock & $w_{opt}$ & $f(w_{opt})$ & $n_{max}$ & Number of tests & Number of increases & Number of decreases \\ \hline
GME   &7240            &15.351               &14343            &89042                 &20179                     &10933                     \\ \hline
BBBY  & 20673      & 25.432         & 22596      & 64582           & 3789                & 11354                \\ \hline
AMC   & 8060       & 14.062         & 16343      & 89855           & 6992                & 12003                     \\ \hline
AAPL  &21515            &24.064               &33849            &104809                 &24966                     &7191                     \\ \hline
CRM  &27647            &5.027              &44287            &88858                 &8204                     & 2058                    \\ \hline
MSFT  &35462            &8.602               &41378            &76696                 &24200                     &11                     \\ \hline
\end{tabular}
\caption*{$f(w_{opt})$ is calculated on the training set. The number of tests is the amount of adjacent time intervals with lengths $w_{opt}$ in the testing set. The increases and decreases are statistically significant changes in entropy between two adjacent intervals.}
\label{Table:2}
\end{table}

In all six cases, entropy still exhibits time-varying behavior. Compared to the previous setting, where the training set is one year, the number of statistically significant changes in entropy for the stocks GME and AAPL increases. Also, in all cases, the maximum value of the objective function, indicating how much the entropies of two adjacent intervals differ, increases.

There are two sequences of statistically significant decreases in entropy value for the stock GME. They start from the entropies calculated at the periods [03.08.20 12:31 to 22.09.20 13:50] and [29.12.20 11:23 to 27.01.21 13:15]. For the stock AMC\footnote{Non-zero returns in the first three quarters of 2019 are not enough to find the intraday volatility pattern for all minutes for the stock AMC. This generates missing values in the filtered return time series of the year 2019.}, there are three series of low entropy values. They start from the entropies calculated at the periods [25.03.20 12:31 to 13.05.20 14:17], [30.09.20 14:21 to 13.11.20 13:31], and [05.05.21 11:58 to 04.06.21 14:51]. For both stocks, the last time intervals correspond to a sharp increase in the prices. For the stock BBBY, the entropy becomes statistically significantly low starting from the interval [07.08.20 11:10 to 18.11.20 11:16]. Thus, entropy was low at the time of the rapid growth in the price and trading volumes.

\clearpage

We can conclude that a possible change in the structure of data regularities over time is not the cause of the changes in the values of entropy. As in the previous section, the entropies of IT companies do not drop as low as in the case of meme stocks. Therefore, there is more predictability in the prices of meme stocks than in the prices of IT companies.
\section{Conclusions and discussion}\label{Conclusions}
We introduced a novel procedure of hypothesis testing to determine if two discrete sequences with the same alphabet have different entropy values. We use it to detect changes in the entropy value of a sequence. Since entropy can change over time, we cannot choose the length of the interval for measuring entropy as long as possible. For this reason, we have introduced three contributions. First, we have found the approximation of entropy's variance that is used in hypothesis testing. Our formula is more precise regarding the length of the interval compared to the formulas proposed by Basharin \cite{basharin59} and Harris \cite{Harris75}. Second, we have introduced an unbiased estimation of the approximation of the variance so that we can evaluate the variance from a sequence. Finally, we have proposed the method for finding an optimal length of the interval. We have shown that this method is suitable for determining how to split a sample into two time series that display statistically different values of Shannon entropy.

We applied the novel method to the return time series of meme stocks. We have found time intervals when entropy is statistically low with respect to other periods, thus signaling a high predictability of the price pattern. In particular, we have focused on three meme stocks, then comparing them with three more standard IT companies, namely GME, BBBY, AMC, AAPL, CRM, and MSFT. We have found that entropy changes over time for all six stocks in a period between January 2020 and July 2021. We can also say that the entropy was time-varying in the year 2019 for all stocks except CRM.

Low entropy values found for each stock are indicative of the predictability of return time series. We filtered data regularities in two ways, and therefore, we believe that the entropy values obtained are related to the measure of market inefficiency. The deviation of the U.S. stock market from efficiency is also detected in other articles, e.g. \cite{Alvarez21,Giglio_2008,Molgedey00}.

From the entropy plots, we notice that entropy for meme stocks falls lower than that for IT stocks. This indicates the existence of the periods of high predictability for meme stocks compared to IT stocks. For the GME stock, a low level of entropy was identified prior to the price spike. Moreover, a fall in entropy corresponds more to the growth in trading volumes than to the increase in the stock price according to the both series of low entropy values. For the AMC stock, entropy was low when the price went up in January 2021. Entropy fell again before the price attained its maximum in June 2021. For the BBBY stock, the period from December 2020 to the end of January 2021 is also characterized by low entropy, although it was discovered late enough to be considered a price increase warning. The low entropy values found for the BBBY stock are more likely to correspond first to the price drop in February 2021 and then to another spike in June 2021.

In the case of GameStop and AMC Entertainment Holdings, growth in prices and trading volumes is characterized by a drop in the entropy value. Interestingly, such a drop occurs before the boom observed in January 2021, already in the late December 2020. That is, some regularity pattern in the price dynamics, that appeared before all the news spread the market, leads to a statistically significant signal of market inefficiency. Given the observed timing, such a signal can also be interpreted as an early-warning of turmoil period for the stock.

Today, financial markets are inherently high-dimensional due to the plethora of instruments composing the portfolios of investors. At the same time, they are highly challenging to monitor, displaying more and more complex cycles, booms and bursts of prices, economic bubbles and so on, all of them representing severe risk factors for the portfolios. In this high-dimensional and complex context, the existence of online early-warnings of market inefficiency is key. In fact, such signals allow to anticipate the periods of turmoil, thus covering or, at least, mitigating the portfolio risk associated with such events, with potential stabilizing effects for the whole market.


\begin{appendices}
\section{List of propositions}
\label{List of propositions}
For the proof of Theorem 1 we need five propositions below.
\begin{proposition}
\label{proposition:1}
Central moments of multinomial distribution $f^M(p_1,p_2,n)$ divided by $n$ (/n) can be defined recursively using the formulas below.
\begin{equation}
\label{eq:prop5}
\begin{split}
\mu_{1,0}&=0, \mu_{1,1}=-\frac{p_1p_2}{n}\\
\mu_{m+1,k}&=\frac{p_1}{n}\left[(1-p_1)\frac{d}{dp_1}\mu_{m,k}-p_2\frac{d}{dp_2}\mu_{m,k}+(1-p_1)m\mu_{m-1,k}-p_2k\mu_{m,k-1}\right]
\end{split}
\end{equation}
\end{proposition}
Proof:
\begin{equation*}
  \mu^M_{m,k}(p_1,p_2,n)=\sum_{x_1\ge 0,x_2\ge 0, x_1+x_2\le n}(x_1-np_1)^m (x_2-np_2)^k \frac{n!}{x_1!x_2!(n-x_1-x_2)!}p_1^{x_1}p_2^{x_2}q^{n-x_1-x_2}
\end{equation*}
where $\mu^M_{m,k}$ is the (m,k)-central moment of the multinomial distribution and $q=1-p_1-p_2$. We can show that
\begin{equation*}
    \begin{split}
        \frac{d}{dp_1}\mu^M_{m,k}&=-nm\mu^M_{m-1,k}+\frac{1-p_2}{p_1q}\mu^M_{m+1,k}+\frac{1}{q}\mu^M_{m,k+1}\\
        \frac{d}{dp_2}\mu^M_{m,k}&=-nk\mu^M_{m,k-1}+\frac{1-p_1}{p_2q}\mu^M_{m,k+1}+\frac{1}{q}\mu^M_{m+1,k}
    \end{split}
\end{equation*} Solving the system for $\mu^M_{m+1,k}$, we get that
\begin{equation*}
\mu^M_{m+1,k}=p_1\left[(1-p_1)\frac{d}{dp_1}\mu^M_{m,k}-p_2\frac{d}{dp_2}\mu^M_{m,k}+(1-p_1)mn\mu^M_{m-1,k}-p_2kn\mu^M_{m,k-1}\right]
\end{equation*}
Taking into account that $\mu^M_{m,k}=n^{m+k}\mu_{m,k}$, we obtain the result
\begin{equation*}
\mu_{m+1,k}=\frac{p_1}{n}\left[(1-p_1)\frac{d}{dp_1}\mu_{m,k}-p_2\frac{d}{dp_2}\mu_{m,k}+(1-p_1)m\mu_{m-1,k}-p_2k\mu_{m,k-1}\right]
\end{equation*}
and  by symmetry
\begin{equation*}
\mu_{m,k+1}=\frac{p_2}{n}\left[(1-p_2)\frac{d}{dp_2}\mu_{m,k}-p_1\frac{d}{dp_1}\mu_{m,k}+(1-p_2)k\mu_{m,k-1}-p_1m\mu_{m-1,k}\right]
\end{equation*}
\begin{proposition}
Central moments of the binomial distribution $B(p,n)/n$ can be defined recursively using the formulas below.
\begin{equation}
\label{eq:moments}
    \begin{split}
    \mu_0&=1\\
    \mu_1&=0\\
    \mu_{m+1}&=\frac{p(1-p)}{n}\left[m\mu_{m-1}+\frac{d}{dp}\mu_{m}\right]
     \end{split}
\end{equation}
\end{proposition}
This is a special case of the previous proposition where $p_2=k=0$. It is known as the Renovsky formula \cite{Riordan37}.
\begin{proposition}
\label{prop: plnp}
\begin{equation}
\label{eq:prop1}
    E(\hat{p}\ln{\hat{p}})=p\ln{p}+\sum_{m=2}^{\infty}\frac{(-1)^{m}}{m(m-1)p^{m-1}}\mu_m
\end{equation}
where $\hat{p}\sim B(p,n)/n$ and $\mu_m$ is its central m-moment.
\end{proposition}
The result of Proposition \ref{prop: plnp} was obtained in \cite{basharin59}. It is derived by using the Taylor expansion around $p$.
\begin{equation*}
\begin{split}
    \hat{p}\ln{\hat{p}}=p\ln{p}+(1+\ln{p})(\hat{p}-p)+\sum_{m=2}^{\infty}\frac{(-1)^{m}}{m(m-1)p^{m-1}}(\hat{p}-p)^{m}
\end{split}
\end{equation*}
Therefore,
\begin{equation*}
    E[\hat{p}\ln{\hat{p}}]=p\ln{p}+\sum_{m=2}^{\infty}\frac{(-1)^{m}}{m(m-1)p^{m-1}}\mu_m.
\end{equation*}
Binomial distribution $B(p,n)$ divided by $n$ has the mean $p$ and the variance $\frac{p(1-p)}{n}$.
\begin{proposition}
Let $\hat{p}\sim B(p,n)/n$. Then,
\label{proposition:4}
\begin{equation}
\label{eq:prop2}
\begin{split}
&E\left(\hat{p}^2\ln^2(\hat{p})\right)=p^2\ln^2(p)+(\ln^2{p}+3\ln{p}+1)\mu_2+\\
&+4\sum_{m=1}^{\infty}(-1)^{m+1}\left[\ln{p}-S_{m-1}+\frac{3}{2}\right]\frac{\mu_{m+2}}{{m(m+1)(m+2)p^{m}}}
\end{split}
\end{equation}
where $S_m=\sum_{k=1}^m\frac{1}{k}$.
\end{proposition}

Proof: We consider the Taylor expansion of $\hat{p}^2\ln^2(\hat{p})$.
\begin{equation*}
\begin{split}
&\hat{p}^2\ln^2(\hat{p})=p^2\ln^2(p)+2p\ln{p}(\ln{p}+1)(\hat{p}-p)+(\ln^2{p}+3\ln{p}+1)(\hat{p}-p)^2+\\
&+4\sum_{m=1}^{\infty}(-1)^{m+1}\left[\ln{p}-S_{m-1}+\frac{3}{2}\right]\frac{(\hat{p}-p)^{m+2}}{{m(m+1)(m+2)p^{m}}}
\end{split}
\end{equation*}
This expression can be obtained by noticing that derivatives of $p^2\ln^2{(p)}$ starting from the third take the form

$$\frac{a_m\ln{p}+b_m}{p^m},$$ 

where $a_{m+1}=-ma_{m}$; $mb_m+b_{m+1}=a_m$ with $a_1=4$; $b_1=6$. The solution of the system is $a_m=4(-1)^{m+1}(m-1)!$ and $b_m=4(-1)^m(m-1)!(S_{m-1}-\frac{3}{2})$. The solution is unique because of the uniqueness of the Taylor series. Taking the expected value, we get the result.
\begin{proposition}
Let $\hat{p}_1, \hat{p}_2\sim f^M(p_1,p_2,n)/n$. Then, 
\begin{equation}
\label{eq:prop3}
\begin{split}
&E(\hat{p}_1\ln{\hat{p}_1}\hat{p}_2\ln{\hat{p}_2})=p_1p_2\ln{p_1}\ln{p_2}+(\ln{p_1}+1)(\ln{p_2}+1)\mu_{1,1}+\\
&+\sum_{m=2}^{\infty}\frac{(-1)^{m}}{m(m-1)}\left[p_1\ln{p_1}\frac{1}{p_2^{m-1}}\mu_{0,m}+p_2\ln{p_2}\frac{1}{p_1^{m-1}}\mu_{m,0}\right]+\\
&+\sum_{m=2}^{\infty}\frac{(-1)^{m}}{m(m-1)}\left[(\ln{p_1}+1)\frac{1}{p_2^{m-1}}\mu_{1,m}+(\ln{p_2}+1)\frac{1}{p_1^{m-1}}\mu_{m,1}\right]+\\
&+\sum_{m=2}^{\infty}\sum_{k=2}^{\infty}\frac{(-1)^{m+k}}{m(m-1)k(k-1)p_1^{m-1}p_2^{k-1}}\mu_{m,k}
    \end{split}
\end{equation}
where $\mu_{m,k}$ are $(m,k)$-central moments of $f^M(p_1,p_2,n)/n$.
\end{proposition}
Proof:
\begin{equation*}
\begin{split}
&\hat{p}_1\ln{\hat{p}_1}\hat{p}_2\ln{\hat{p}_2}=\sum_{m=0}^{\infty}\frac{(\hat{p}_1-p_1)^m}{m!}\frac{d^m}{dp_1^m}{(p_1\ln{p_1})}\sum_{k=0}^{\infty}\frac{(\hat{p}_2-p_2)^k}{k!}\frac{d^k}{dp_2^k}{(p_2\ln{p_2})}=\\
&=p_1p_2\ln{p_1}\ln{p_2}+p_1\ln{p_1}(\ln{p_2}+1)(\hat{p}_2-p_2)+p_2\ln{p_2}(\ln{p_1}+1)(\hat{p}_1-p_1)+\\
&+p_1\ln{p_1}\sum_{k=2}^{\infty}\frac{(-1)^{k}}{k(k-1)p_2^{k-1}}(\hat{p}_2-p_2)^{k}+p_2\ln{p_2}\sum_{m=2}^{\infty}\frac{(-1)^{m}}{m(m-1)p_1^{m-1}}(\hat{p}_1-p_1)^{m}+\\
&+(\ln{p_1}+1)(\hat{p}_1-p_1)\sum_{k=2}^{\infty}\frac{(-1)^{k}}{k(k-1)p_2^{k-1}}(\hat{p}_2-p_2)^{k}+(\ln{p_2}+1)(\hat{p}_2-p_2)\sum_{m=2}^{\infty}\frac{(-1)^{m}}{m(m-1)p_1^{m-1}}(\hat{p}_1-p_1)^{m}+\\
&+(\ln{p_1}+1)(\ln{p_2}+1)(\hat{p}_1-p_1)(\hat{p}_2-p_2)+\sum_{m=2}^{\infty}\sum_{k=2}^{\infty}\frac{(-1)^{m+k}}{m(m-1)k(k-1)p_1^{m-1}p_2^{k-1}}(\hat{p}_1-p_1)^{m}(\hat{p}_2-p_2)^{k}
    \end{split}
\end{equation*}
Therefore,
\begin{equation*}
\begin{split}
&E(\hat{p}_1\ln{\hat{p}_1}\hat{p}_2\ln{\hat{p}_2})=p_1p_2\ln{p_1}\ln{p_2}+\\
&+p_1\ln{p_1}\sum_{k=2}^{\infty}\frac{(-1)^{k}}{k(k-1)p_2^{k-1}}\mu_{0,k}+p_2\ln{p_2}\sum_{m=2}^{\infty}\frac{(-1)^{m}}{m(m-1)p_1^{m-1}}\mu_{m,0}+\\
&+(\ln{p_1}+1)\sum_{k=2}^{\infty}\frac{(-1)^{k}}{k(k-1)p_2^{k-1}}\mu_{1,k}+(\ln{p_2}+1)\sum_{m=2}^{\infty}\frac{(-1)^{m}}{m(m-1)p_1^{m-1}}\mu_{m,1}+\\
&+(\ln{p_1}+1)(\ln{p_1}+1)\mu_{1,1}+\sum_{m=2}^{\infty}\sum_{k=2}^{\infty}\frac{(-1)^{m+k}}{m(m-1)k(k-1)p_1^{m-1}p_2^{k-1}}\mu_{m,k}
    \end{split}
\end{equation*}
\section{Proof of Theorem 1}
\label{Proof of the Theorem 1}
\begin{equation*}
    \begin{split}
    Var(\hat{H})&=E(\hat{H}^2)-E(\hat{H})^2\\
    &=\sum_{j=0}^{M-1}E({\hat{p}_j^2\ln^2{\hat{p}_j}})+\sum_{j=0}^{M-1}\sum_{i=0, i\neq j}^{M-1}E({\hat{p}_j\ln{\hat{p}_j}\hat{p_i}\ln{\hat{p}_i}})-E(\hat{H})^2
         \end{split}
\end{equation*}
For calculations we need all moments of orders $n^{-1}$, $n^{-2}$, $n^{-3}$ obtained using Equations \ref{eq:prop5} and \ref{eq:moments}.
\begin{equation*}
    \begin{split}
    \mu_2&=\frac{p(1-p)}{n}\\
    \mu_3&=\frac{p(1-p)(1-2p)}{n^2}\\
    \mu_{4}&=\frac{3p^2(1-p)^2}{n^2}+\frac{p(1-p)-6p^2(1-p)^2}{n^3}\\
    \mu_{5}&=\frac{10p^2(1-p)^2(1-2p)}{n^3}+O(n^{-4})\\
     \mu_{6}&=\frac{15p^3(1-p)^3}{n^3}+O(n^{-4})
     \end{split}
\end{equation*}
\begin{equation*}
    \begin{split}
    \mu_{2,1}&=-\frac{p_1p_2(1-2p_1)}{n^2}\\
    \mu_{3,1}&=-\frac{3p_1^2(1-p_1)p_2}{n^2}+\frac{6p_1^2(1-p_1)p_2-p_1p_2}{n^3}\\
    \mu_{4,1}&=-\frac{10p_1^2(1-p_1)(1-2p_1)p_2}{n^3}+O(n^{-4})\\
    \mu_{5,1}&=-\frac{15p_1^3(1-p_1)^2p_2}{n^3}+O(n^{-4})
     \end{split}
\end{equation*}
\begin{equation*}
    \begin{split}
\mu_{2,2}&=\frac{p_1p_2(1-p_1)(1-p_2)+2p_1^2p_2^2}{n^2}+\frac{p_1p_2-2p_1p_2(1-p_1)(1-p_2)-4p_1^2p_2^2}{n^3}\\
\mu_{3,2}&=\frac{10p_1^2p_2^2(1-2p_1)+p_1p_2(1-p_1-p_2)(1-5p_1)}{n^3}\\
\mu_{3,3}&=-\frac{9p_1^2p_2^2(1-p_1)(1-p_2)+6p_1^3p_2^3}{n^3}+O(n^{-4})\\
\mu_{4,2}&=\frac{3p_1^2(1-p_1)p_2((1-p_1)(1-p_2)+4p_1p_2)}{n^3}+O(n^{-4})
     \end{split}
\end{equation*}
Moments with $m+k\le4$ coincide with results obtained in \cite{Harris75,Ouimet20}. After summing up $E[\hat{p_j}\ln{\hat{p_j}}]$ in Eq.~\ref{eq:prop1} for all $j$, the expression becomes
\begin{equation}
\label{eq:Entropy}
E(\hat{H})=-E(\sum_j\hat{p}_j\ln(\hat{p}_j))=H-\frac{M-1}{2n}+\frac{1}{12n^2}\left(1-\sum_{j=0}^{M-1}\frac{1}{p_j}\right)+\frac{1}{12n^3}\sum_{j=0}^{M-1}\left(\frac{1}{p_j}-\frac{1}{p_j^2}\right)+O(n^{-4})
\end{equation}

where $H=-\sum_jp_j\ln(p_j)$, $\mu_2,\mu_3,\mu_4,\mu_5,\mu_6$ are used. Similar estimates of the bias of entropy estimation were obtained in other works, see, e.g. \cite{Harris75, Schurmann96, Victor00}.
\begin{equation*}
    \begin{split}
    E(\hat{H})^2&=H^2+\frac{1}{n}\left[-(M-1)H\right]+\frac{1}{n^2}\left[\frac{M^2}{4}-\frac{M}{2}+\frac{1}{4}+\frac{H}{6}\left(1-\sum_{j=0}^{M-1}\frac{1}{p_j}\right)\right]+\\
    &+\frac{1}{n^3}\left[\frac{H}{6}\sum_{j=0}^{M-1}\left(\frac{1}{p_j}-\frac{1}{p_j^2}\right)-\frac{M-1}{12}\left(1-\sum_{j=0}^{M-1}\frac{1}{p_j}\right)\right]
         \end{split}
\end{equation*}
The approximation of the second moment of $\hat{p}\ln(\hat{p})$ from Eq.~\ref{eq:prop2} is
\begin{equation*}
\begin{split}
&E(\hat{p}^2\ln^2(\hat{p}))=p^2\ln^2(p)+\frac{1}{n}\left(\ln^2{p}+3\ln{p}+1\right)p(1-p)+\\
&+\frac{1}{n^2}\left[\left(\frac{5}{6}p^2-p+\frac{1}{6}\right)\ln{p}+\frac{7}{4}p^2-\frac{5}{2}p+\frac{3}{4}\right]+\\
&+\frac{1}{n^3}\left[\frac{1}{6}\left(p^2-p\right)\ln{p}+\frac{p^2}{3}-\frac{p}{2}+\frac{1}{12}+\frac{1}{12p}\right]+O(n^{-4})
\end{split}
\end{equation*}
The approximation of the covariances from Eq.~\ref{eq:prop3} is
\begin{equation*}
\begin{split}
&E(\hat{p}_1\ln{\hat{p}_1}\hat{p}_2\ln{\hat{p}_2})=\\
&=p_1p_2\ln{p_1}\ln{p_2}+\frac{1}{n}\left[-(\ln{p_1}+1)(\ln{p_2}+1)p_1p_2+\frac{1}{2}\left(p_1\ln{p_1}(1-p_2)+p_2\ln{p_2}(1-p_1)\right)\right]+\\
&+\frac{1}{n^2}\left[\frac{5}{12}p_1p_2(\ln{p_1}+\ln{p_2})+\frac{1}{12}\left(\frac{p_1}{p_2}\ln{p_1}+\frac{p_2}{p_1}\ln{p_2}\right)+\frac{1}{4}(1+7p_1p_2-p_1-p_2)\right]+\\
&+\frac{1}{n^3}\left[\frac{1}{12}p_1\ln{p_1}\left(p_2+\frac{1}{p_2^2}\right)+\frac{1}{12}p_2\ln{p_2}\left(p_1+\frac{1}{p_1^2}\right)+\frac{1}{3}p_1p_2+\frac{1}{24}\left(\frac{p_1}{p_2}+\frac{p_2}{p_1}+\frac{1}{p_1}+\frac{1}{p_2}-p_1-p_2\right)\right]+\\
&+O(n^{-4})
    \end{split}
\end{equation*}
Summing up for all indexes $j,i$ of the second moments and covariances, we get that
\begin{equation*}
\begin{split}
E(\hat{H}^2)&=H^2+\frac{1}{n}\left[-H^2+\sum_j p_j\ln^2(p_j)-(M-1)H\right]+
\\&+\frac{1}{n^2}\left[\frac{H}{6}\left(1-\sum_j\frac{1}{p_j}\right)+\frac{1}{4}M^2-\frac{1}{4}\right]+
\\&+\frac{1}{n^3}\left[\frac{M}{12}\sum_j\frac{1}{p_j}+\frac{1}{12}\sum_j\frac{1}{p_j}-\frac{1}{12}-\frac{M}{12}-\frac{1}{6}H\sum_j\frac{1}{p_j^2}-\frac{1}{6}\sum_j\frac{\ln{p_j}}{p_j}\right]+O(n^{-4})
    \end{split}
\end{equation*}
Therefore,
\begin{equation*}
Var(\hat{H})=\frac{1}{n}\left[-H^2+\sum_j p_j\ln^2(p_j)\right]+\frac{1}{n^2}\left[\frac{M}{2}-\frac{1}{2}\right]+\frac{1}{6n^3}\left[(1-H)\sum_j\frac{1}{p_j}-\sum_j{\frac{\ln{p_j}}{p_j}}-1\right]+O(n^{-4})
\end{equation*}
\section{Proof of Theorem 2}
\label{Proof of Theorem 2}
We introduce a random variable $\hat{Var}$.
\begin{equation*}
    \begin{split}
\hat{Var}&=\frac{1}{n}(\sum_j \hat{p}_j\ln^2{\hat{p}_j}-\hat{H}^2)+\frac{1}{n^2}\left[\sum_j \hat{p}_j\ln^2{\hat{p}_j}-\hat{H}^2-M\hat{H}-\sum_j{\ln{\hat{p}}}-\frac{M}{2}+\frac{1}{2}\right]+\\
&+\frac{1}{n^3}\left[\sum_j \hat{p}_j\ln^2{\hat{p}_j}-\hat{H}^2-M\hat{H}-\sum_j{\ln{\hat{p}_j}}-\frac{\hat{H}}{3}\sum_j{\frac{1}{\hat{p}_j}}-\frac{1}{3}\sum_j{\frac{\ln{\hat{p}_j}}{\hat{p}_j}}-\frac{1}{12}\sum_j{\frac{1}{\hat{p}_j}}-\frac{M^2}{4}-\frac{M}{2}+\frac{5}{6}\right]
    \end{split}
\end{equation*}
From the proof of Theorem 1 we know that
\begin{equation*}
\begin{split}
E(\hat{H}^2)&=H^2+\frac{1}{n}\left[-H^2+\sum_j p_j\ln^2(p_j)-MH+H\right]+
\\&+\frac{1}{n^2}\left[\frac{H}{6}-\frac{H}{6}\sum_j\frac{1}{p_j}+\frac{1}{4}M^2-\frac{1}{4}\right]+O(n^{-3})
    \end{split}
\end{equation*}
and
\begin{equation*}
E(M\hat{H})=MH+\frac{M-M^2}{2n}+O(n^{-2}).
\end{equation*}
We can show using Taylor series and moments $\mu_2, \mu_3, \mu_4$ that
\begin{equation*}
\begin{split}
E\left(\sum_j\hat{p}_j\ln^2\hat{p}_j\right)&=\sum_j p_j\ln^2{p_j}+\frac{1}{n}\left(\sum_j\ln{p_j}+H+M-1\right)+\\
&+\frac{1}{n^2}\left(\frac{1}{6}\sum_j\frac{\ln{p_j}}{p_j}+\frac{M}{2}-\frac{1}{4}\sum_j\frac{1}{p_j}+\frac{H}{6}-\frac{1}{4}\right)+O(n^{-3})
\end{split}
\end{equation*}
and
\begin{equation*}
E\left(\sum_j\ln{\hat{p}_j}\right)=\sum_j\ln{p_j}+\frac{1}{n}\left(\frac{M}{2}-\frac{1}{2}\sum_j\frac{1}{p_j}\right)+O(n^{-2}).
\end{equation*}

We get the result by using the equation

\begin{equation*}
E\left(-\frac{\hat{H}}{3}\sum_j{\frac{1}{\hat{p}_j}}-\frac{1}{3}\sum_j{\frac{\ln{\hat{p}_j}}{\hat{p}_j}}-\frac{1}{12}\sum_j{\frac{1}{\hat{p}_j}}\right)=-\frac{H}{3}\sum_j{\frac{1}{p_j}}-\frac{1}{3}\sum_j{\frac{\ln{p_j}}{p_j}}-\frac{1}{12}\sum_j{\frac{1}{p_j}}+O(n^{-1})
\end{equation*}

and substituting all equations in the formula for $E(\hat{Var})$.

\begin{equation*}
E(\hat{Var})=\frac{1}{n}\left[-H^2+\sum_j p_j\ln^2(p_j)\right]+\frac{1}{n^2}\left[\frac{M}{2}-\frac{1}{2}\right]+\frac{1}{6n^3}\left[(1-H)\sum_j\frac{1}{p_j}-\sum_j{\frac{\ln{p_j}}{p_j}}-1\right]+O(n^{-4})
\end{equation*}
\section{Entropies, prices, volumes of BBBY and AMC stocks}
\label{Plots}
\begin{figure}[ht]%
\centering
    \includegraphics[width=8.5 cm]{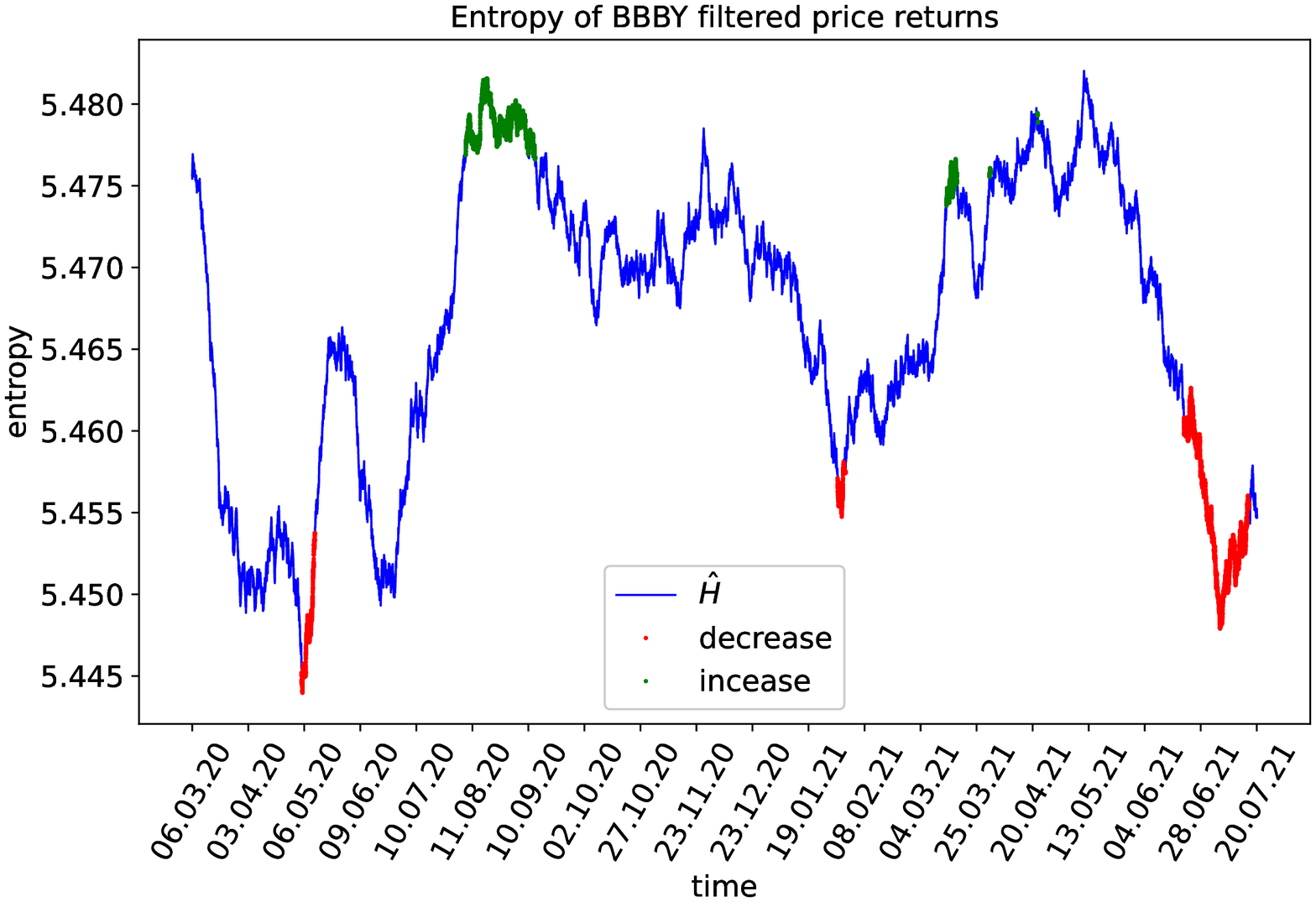}
    \qquad
    \includegraphics[width=8.5 cm]{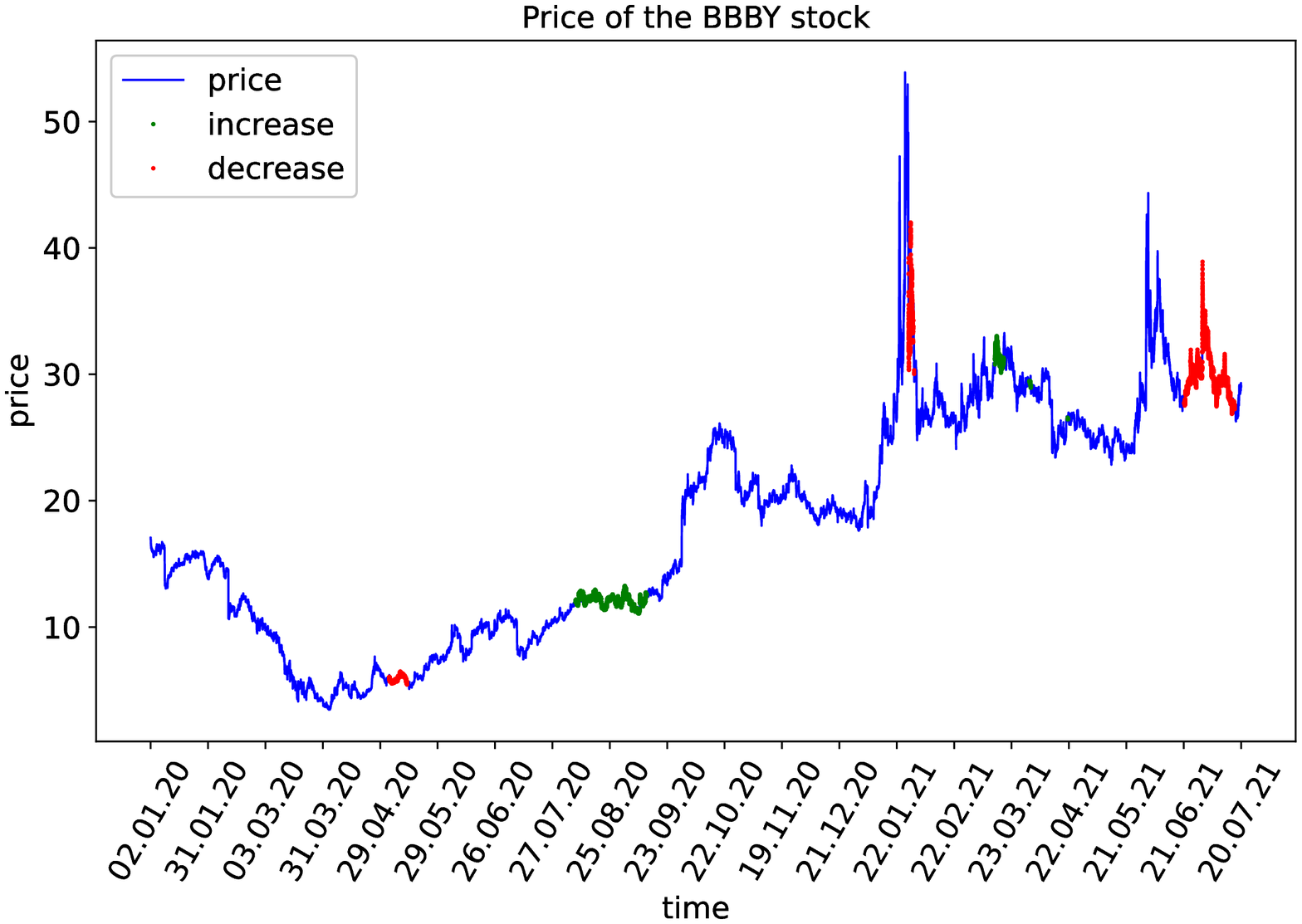}
    \qquad
    \includegraphics[width=8.5 cm]{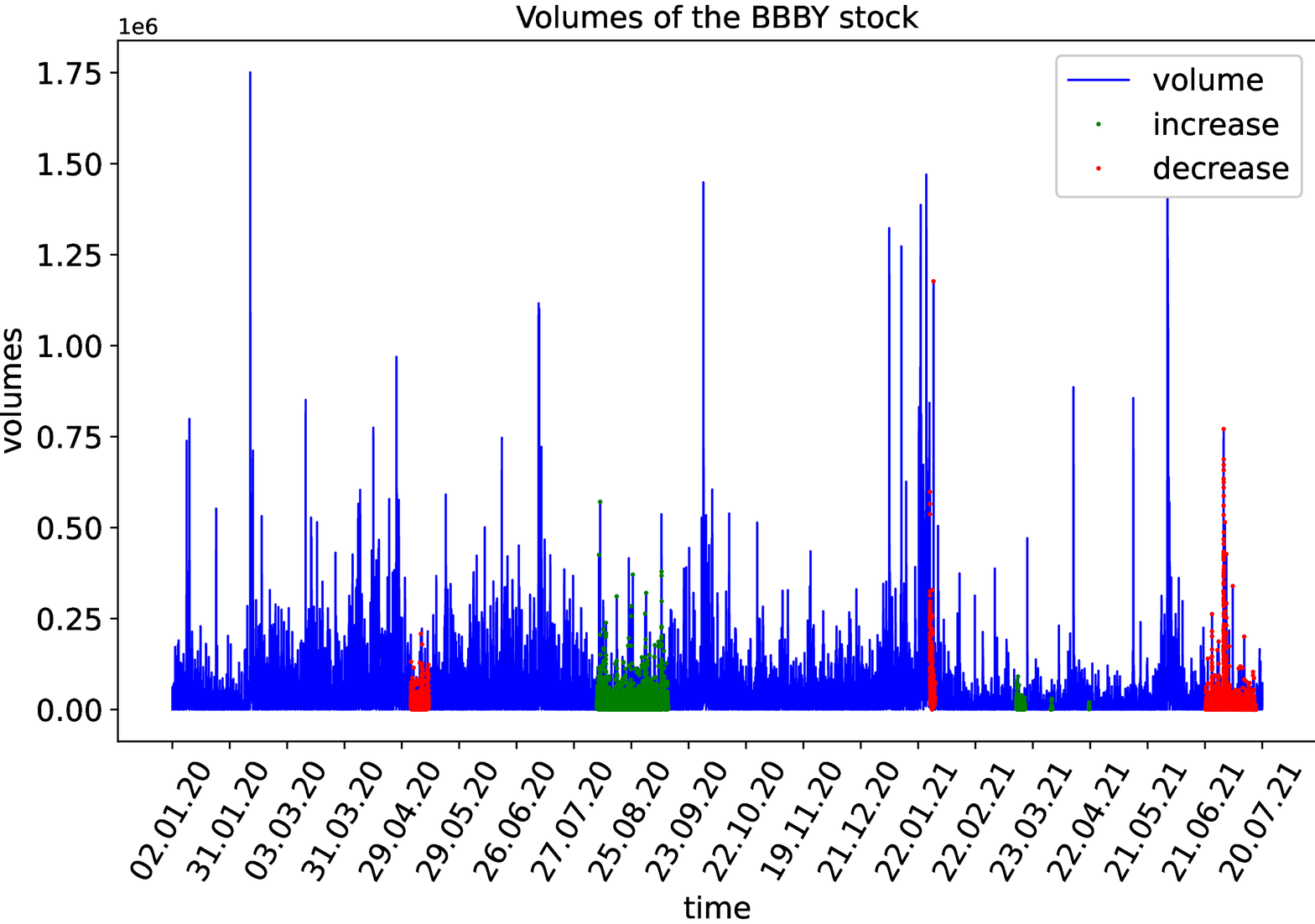}
    \caption{Entropy, price, volume of the BBBY stock. Dots correspond to statistically significant changes in entropy.}
    \label{BBBY stock}%
\end{figure}

\begin{figure}[ht]%
\centering
    \includegraphics[width=8.5 cm]{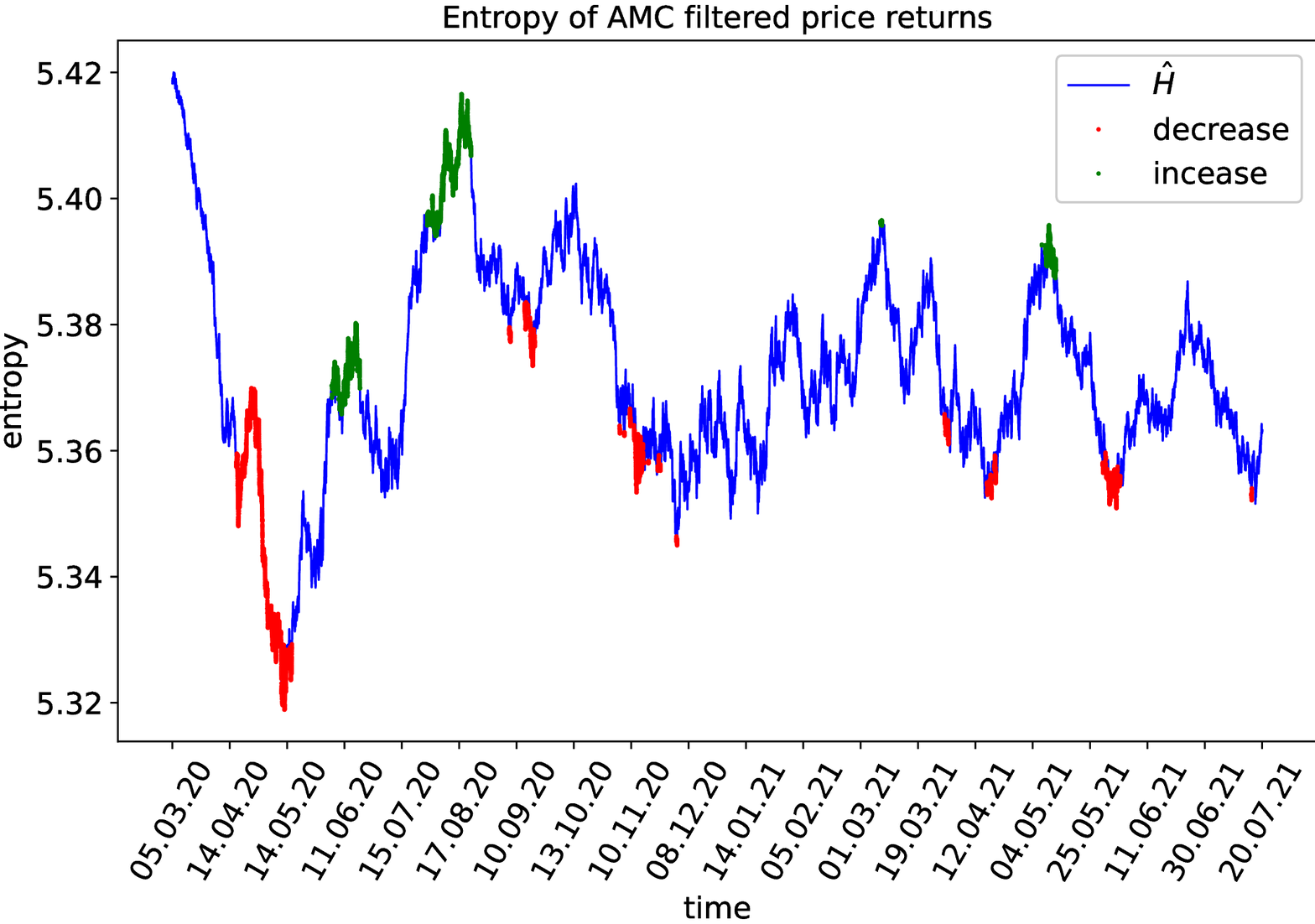}
    \qquad
    \includegraphics[width=8.5 cm]{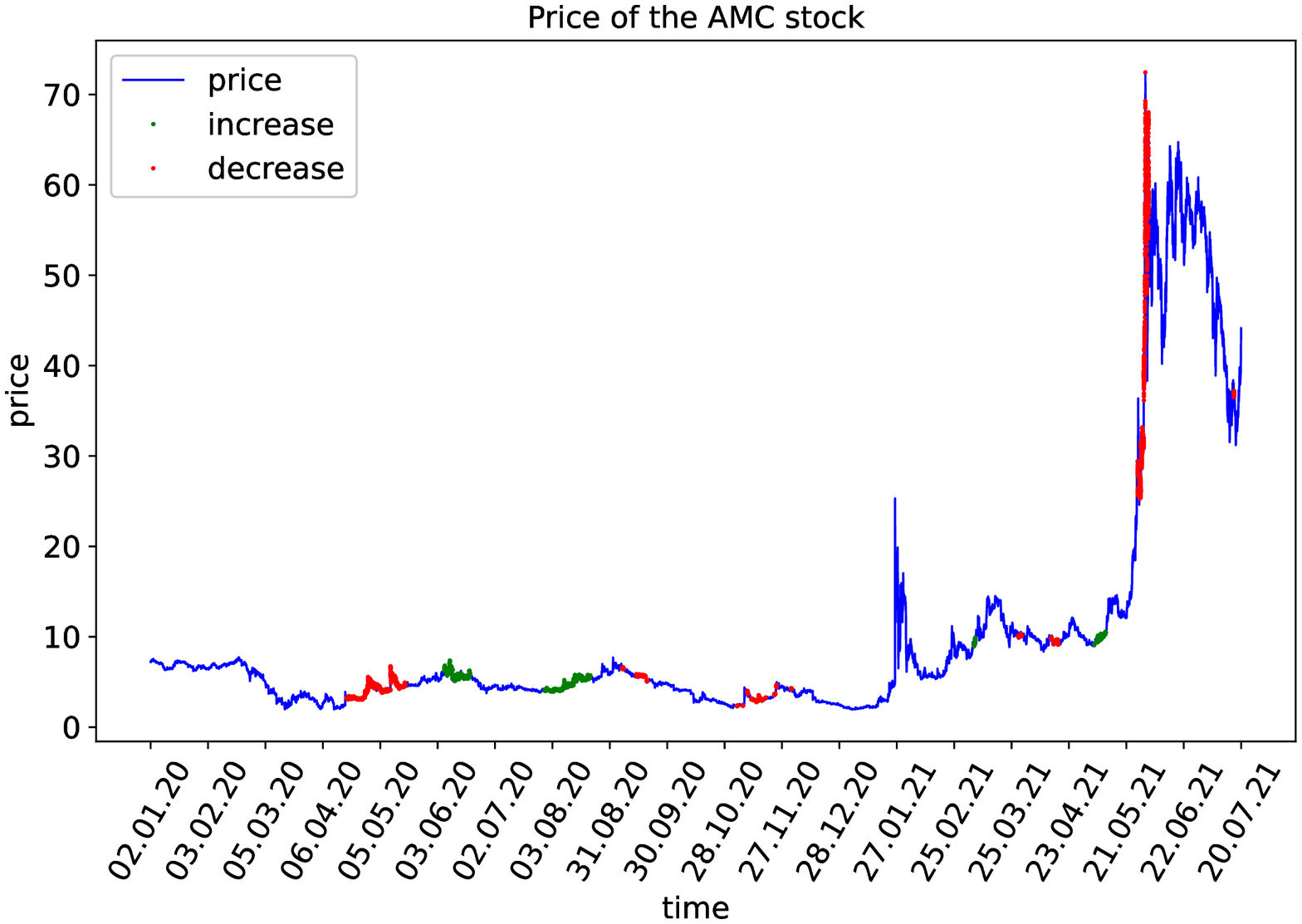}
    \qquad
    \includegraphics[width=8.5 cm]{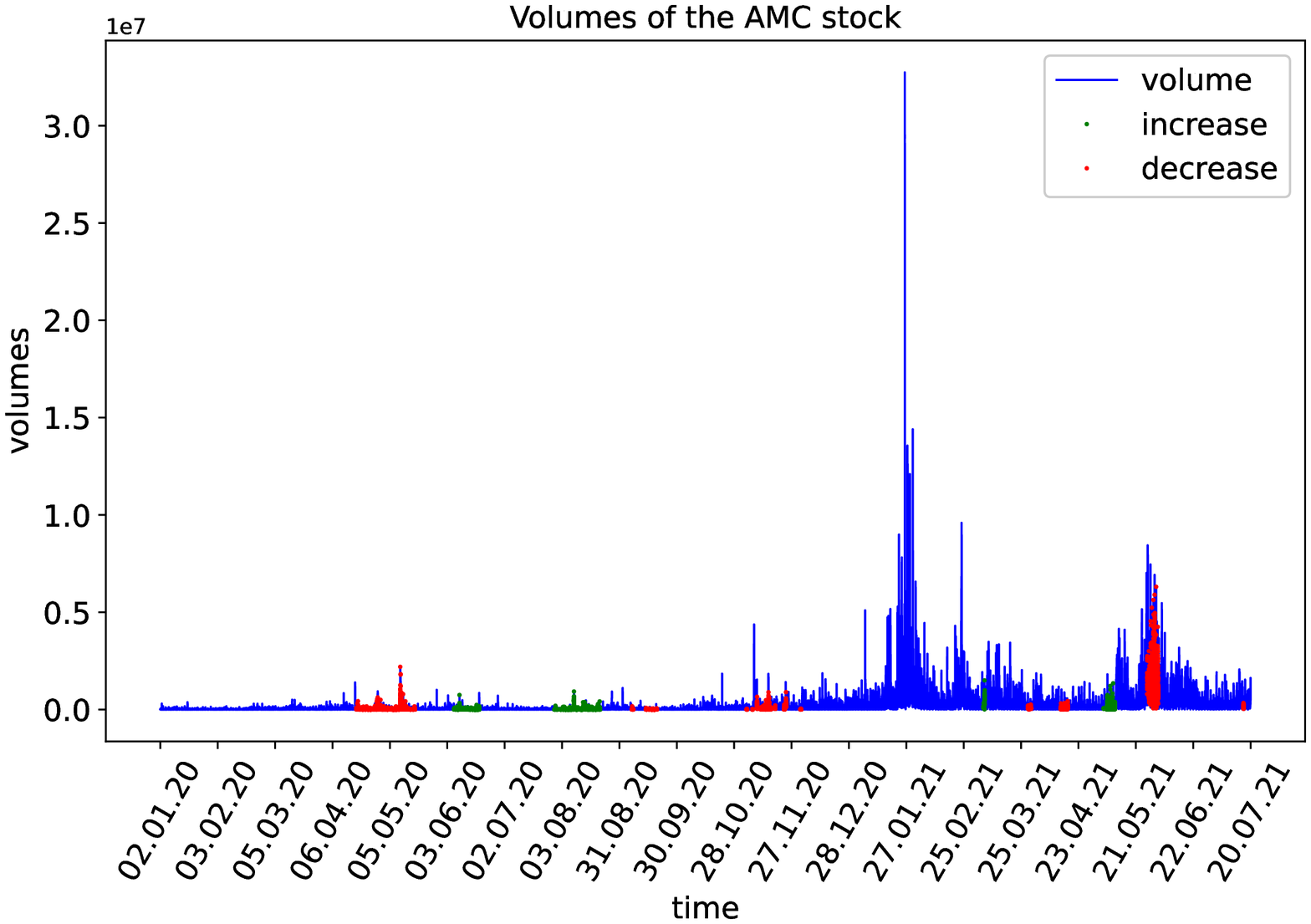}
    \caption{Entropy, price, volume of the AMC stock. Dots correspond to statistically significant changes in entropy.}
    \label{AMC stock}%
\end{figure}

Figures \ref{BBBY stock} and \ref{AMC stock} show estimated entropies, prices, and trading volumes for the stocks BBBY and AMC, respectively.
\clearpage

\end{appendices}
\clearpage
\bibliographystyle{unsrt}
\bibliography{bibfile}

\begin{thebibliography}{10}

\bibitem{Pincus91}
Steve Pincus, Igor Gladstone, and Richard Ehrenkranz.
\newblock A regular statistic for medical data analysis.
\newblock {\em Journal of clinical monitoring}, 7:335--345, 11 1991.

\bibitem{Pincus04}
Steve Pincus and Rudolf Kalman.
\newblock Irregularity, volatility, risk, and financial market time series.
\newblock {\em Proceedings of the National Academy of Sciences of the United
  States of America}, 101:13709--14, 10 2004.

\bibitem{Dong19}
Xinzheng Dong, Chang Chen, Qingshan Geng, Zhixin Cao, Xiaoyan Chen, Jinxiang
  Lin, Yu~Jin, Zhaozhi Zhang, Yan Shi, and Xiaohua~Douglas Zhang.
\newblock An improved method of handling missing values in the analysis of
  sample entropy for continuous monitoring of physiological signals.
\newblock {\em Entropy}, 21(3), 2019.

\bibitem{Pandey15}
Biswajit Pandey and Suman Sarkar.
\newblock {Testing homogeneity in the Sloan Digital Sky Survey Data Release
  Twelve with Shannon entropy}.
\newblock {\em Monthly Notices of the Royal Astronomical Society},
  454(3):2647--2656, 10 2015.

\bibitem{Strait96}
B.J. Strait and T.G. Dewey.
\newblock The shannon information entropy of protein sequences.
\newblock {\em Biophysical Journal}, 71(1):148--155, 1996.

\bibitem{Bezerianos03}
Anastasios Bezerianos, Shanbao Tong, and N.v Thakor.
\newblock Time-dependent entropy estimation of eeg rhythm changes following
  brain ischemia.
\newblock {\em Annals of biomedical engineering}, 31:221--32, 03 2003.

\bibitem{Samuelson}
Paul~Anthony Samuelson.
\newblock Proof that properly anticipated prices fluctuate randomly.
\newblock {\em Ind. manage. rev.}, 6:41--49, 1965.

\bibitem{Fama}
Eugene~F. Fama.
\newblock Efficient capital markets: A review of theory and empirical work.
\newblock {\em J Finance}, 25:383--417, 5 1970.

\bibitem{Barnett00}
William~A. Barnett and Apostolos Serletis.
\newblock Martingales, nonlinearity, and chaos.
\newblock {\em Journal of Economic Dynamics and Control}, 24(5):703--724, 2000.

\bibitem{Agliari18}
Anna Agliari, Ahmad Naimzada, and Nicolò Pecora.
\newblock Boom-bust dynamics in a stock market participation model with
  heterogeneous traders.
\newblock {\em Journal of Economic Dynamics and Control}, 91:458--468, 2018.
\newblock Special Issue in Honour of Prof. Carl Chiarella.

\bibitem{Chan21}
Joshua~C.C. Chan and Caterina Santi.
\newblock Speculative bubbles in present-value models: A bayesian
  markov-switching state space approach.
\newblock {\em Journal of Economic Dynamics and Control}, 127:104101, 2021.

\bibitem{Molgedey00}
L.~Molgedey and W.~Ebeling.
\newblock Local order, entropy and predictability of financial time series.
\newblock {\em Eur. Phys. J. B}, 15:733--737, 2000.

\bibitem{Risso08}
Wiston~Adrián Risso.
\newblock The informational efficiency and the financial crashes.
\newblock {\em J Int Bus Stud}, 22:396--408, 9 2008.

\bibitem{Mensi}
Walid Mensi, Chaker Aloui, Manel Hamdi, and Duc~Khuong Nguyen.
\newblock Crude oil market efficiency: An empirical investigation via the
  shannon entropy.
\newblock {\em écon. intern.}, 129:119--137, 8 2012.

\bibitem{Olbrys}
Joanna Olbryś and Elżbieta Majewska.
\newblock Regularity in stock market indices within turbulence periods: The
  sample entropy approach.
\newblock {\em Entropy}, 24(7), 2022.

\bibitem{susmel2000}
Raul Susmel.
\newblock Switching volatility in private international equity markets.
\newblock {\em International Journal of Finance \& Economics}, 5(4):265--283,
  2000.

\bibitem{Lobo98}
Bento~J. Lobo and David Tufte.
\newblock Exchange rate volatility: Does politics matter?
\newblock {\em Journal of Macroeconomics}, 20(2):351--365, 1998.

\bibitem{Malik05}
Farooq Malik, Bradley~T Ewing, and James~E Payne.
\newblock Measuring volatility persistence in the presence of sudden changes in
  the variance of canadian stock returns.
\newblock {\em Canadian Journal of Economics/Revue canadienne
  d'{\'e}conomique}, 38(3):1037--1056, 2005.

\bibitem{Calcagnile}
Lucio~Maria Calcagnile, Fulvio Corsi, and Stefano Marmi.
\newblock Entropy and efficiency of the etf market.
\newblock {\em Comput. Econ.}, 55:143--184, 1 2020.

\bibitem{Shternshis2}
Andrey Shternshis, Piero Mazzarisi, and Stefano Marmi.
\newblock Efficiency of the moscow stock exchange before 2022.
\newblock {\em Entropy}, 24(9), 2022.

\bibitem{Marton}
Katalin Marton and Paul~C. Shields.
\newblock Entropy and the consistent estimation of joint distributions.
\newblock {\em Ann. Probab.}, 22:960--977, 4 1994.

\bibitem{basharin59}
G.~P. Basharin.
\newblock On a statistical estimate for the entropy of a sequence of
  independent random variables.
\newblock {\em Theory of Probability \& Its Applications}, 4(3):333--336, 1959.

\bibitem{Jabloun}
Antonio Dávalos, Meryem Jabloun, Philippe Ravier, and Olivier Buttelli.
\newblock On the statistical properties of multiscale permutation entropy:
  Characterization of the estimator’s variance.
\newblock {\em Entropy}, 21(5), 2019.

\bibitem{Harris75}
Bernard Harris.
\newblock The statistical estimation of entropy in the non-parametric case.
\newblock Technical report, Wisconsin Univ-Madison Mathematics Research Center,
  1975.

\bibitem{Ricci21}
Leonardo Ricci, Alessio Perinelli, and Michele Castelluzzo.
\newblock Estimating the variance of shannon entropy.
\newblock {\em Phys. Rev. E}, 104:024220, Aug 2021.

\bibitem{MATILLAGARCIA07}
Mariano Matilla-García.
\newblock A non-parametric test for independence based on symbolic dynamics.
\newblock {\em Journal of Economic Dynamics and Control}, 31(12):3889--3903,
  2007.

\bibitem{sec_report}
Staff report on equity and options market structure conditions in early 2021
  (\href{https://www.sec.gov/files/staff-report-equity-options-market-struction-conditions-early-2021.pdf}{https://www.sec.gov/files/staff-report-equity-options-market-struction-conditions-early-2021.pdf}).
\newblock Technical report, U.S. Securities and Exchange Commission, 2021.

\bibitem{Shannon}
C.~E. Shannon.
\newblock A mathematical theory of communication.
\newblock {\em The Bell System Technical Journal}, 27(3):379--423, 1948.

\bibitem{Cont}
R.~Cont.
\newblock Empirical properties of asset returns: stylized facts and statistical
  issues.
\newblock {\em Quantitative Finance}, 1(2):223--236, 2001.

\bibitem{Wood}
Robert~A. Wood, Thomas~H. McInish, and J.~Keith Ord.
\newblock An investigation of transactions data for nyse stocks.
\newblock {\em The Journal of Finance}, 40(3):723--739, 1985.

\bibitem{Zubkov74}
A.~M. Zubkov.
\newblock Limit distributions for a statistical estimate of the entropy.
\newblock {\em Theory of Probability \& Its Applications}, 18(3):611--618,
  1974.

\bibitem{Mathai93}
A.M. Mathai.
\newblock On noncentral generalized laplacianness of quadratic forms in normal
  variables.
\newblock {\em Journal of Multivariate Analysis}, 45(2):239--246, 1993.

\bibitem{Sidak67}
Zbyněk Šidák.
\newblock Rectangular confidence regions for the means of multivariate normal
  distributions.
\newblock {\em Journal of the American Statistical Association},
  62(318):626--633, 1967.

\bibitem{Alvarez21}
Jose Alvarez-Ramirez and Eduardo Rodriguez.
\newblock A singular value decomposition entropy approach for testing stock
  market efficiency.
\newblock {\em Physica A: Statistical Mechanics and its Applications},
  583:126337, 2021.

\bibitem{Giglio_2008}
R.~Giglio, R.~Matsushita, A.~Figueiredo, I.~Gleria, and S.~Da Silva.
\newblock Algorithmic complexity theory and the relative efficiency of
  financial markets.
\newblock {\em {EPL} (Europhysics Letters)}, 84(4):48005, nov 2008.

\bibitem{Riordan37}
John Riordan.
\newblock Moment recurrence relations for binomial, poisson and hypergeometric
  frequency distributions.
\newblock {\em The Annals of Mathematical Statistics}, 8(2):103--111, 1937.

\bibitem{Ouimet20}
Frédéric Ouimet.
\newblock General formulas for the central and non-central moments of the
  multinomial distribution.
\newblock {\em Stats}, 4:18--27, 2021.

\bibitem{Schurmann96}
Thomas Sch{\"u}rmann and Peter Grassberger.
\newblock Entropy estimation of symbol sequences.
\newblock {\em Chaos: An Interdisciplinary Journal of Nonlinear Science},
  6(3):414--427, 1996.

\bibitem{Victor00}
Jonathan~D. Victor.
\newblock Asymptotic bias in information estimates and the exponential (bell)
  polynomials.
\newblock {\em Neural Computation}, 12(12):2797--2804, 2000.

\end{thebibliography}
\end{document}